\documentstyle[12pt,aasms4]{article}

\newcommand{\gsim}{\mbox{$\stackrel{_>}{_\sim}$}} 
\newcommand{\lsim}{\mbox{$\stackrel{_<}{_\sim}$}} 
\newcommand{\etal}{{et~al.}}

\newcommand{\VOT}{\mbox{${\mathrm{VOT}}$}}    
\newcommand{\AFO}{\mbox{$\mathrm{AFO}$}}      
\newcommand{\Vout}{\mbox{$V_{\mathrm{out}}$}} 

\newcommand{\Vmin}{\mbox{$V_{\mathrm{min}}$}} 
\newcommand{\Vmax}{\mbox{$V_{\mathrm{max}}$}} 

\newcommand{\Voutadu}{\mbox{$V_{\mathrm{out}}^{\mathrm{adu}}$}} 

\newcommand{\adu}{\mbox{$\mathrm{adu}$}}   

\newcommand{\Nbits}{\mbox{$N_{\mathrm{bits}}$}}     

\newcommand{\trunc}{\mbox{$\mathrm{trunc}$}}     

\newcommand{\stringIn}{\mbox{$S_{\mathrm{In}}$}}                 
\newcommand{\stringOut}{\mbox{$S_{\mathrm{Out}}$}}               

\newcommand{\SetstringIn}{\mbox{$\{S_{\mathrm{In}}\}$}}          
\newcommand{\SetstringOut}{\mbox{$\{S_{\mathrm{Out}}\}$}}        

\newcommand{\cmpfunc}{\mbox{${\mathcal{F}}_{\mathrm{Comp}}$}}         

\newcommand{\LIn}{\mbox{$\Nbits$}}                                               
\newcommand{\LOut}{\mbox{$N_{\mathrm{bits}}^{\mathrm{out}}$}}                    
\newcommand{\ALOut}{\mbox{$\overline{N_{\mathrm{bits}}^{\mathrm{out}}}$}}        

\newcommand{\Lchunck}{\mbox{$L_{\mathrm{chunck}}$}}                   

\newcommand{\Lu}{\mbox{$L_{\mathrm{u}}$}}                        
\newcommand{\Lc}{\mbox{$L_{\mathrm{c}}$}}                        
\newcommand{\etac}{\mbox{$\eta_{\mathrm{c}}$}}                   

\newcommand{\Cr}{\mbox{$C_{\mathrm{r}}$}}                        
\newcommand{\CrTh}{\mbox{$C_{\mathrm{r}}^{\mathrm{Th}}$}}        

\newcommand{\CrMSB}{\mbox{$C_{\mathrm{r}}^{\mathrm{MSB}}$}}      
\newcommand{\CrLSB}{\mbox{$C_{\mathrm{r}}^{\mathrm{LSB}}$}}      

\newcommand{\CrFit}{\mbox{$C_{\mathrm{r}}^{\mathrm{Fit}}$}}      
\newcommand{\CrOne}{\mbox{$C_{\mathrm{r},1}$}}                   

\newcommand{\SlopeOne}{\mbox{${\mathcal{S}}_{1}$}}                   

\newcommand{\Nc}{\mbox{$N_{\mathrm{c}}$}}                         

\newcommand{\sign}{$\mbox{{\tt{sign}}}$}                          

\newcommand{\xth}{\mbox{$x_{\mathrm{th}}$}}

\newcommand{\NCirc}{\mbox{$N_{\mathrm{circ}}$}}
\newcommand{\aetac}{\mbox{$\overline{\eta}_{\mathrm{{c}}}$}}
\newcommand{\daetac}{\mbox{$\dot{\overline{\eta}}_{\mathrm{{c}}}$}}

\input epsf.sty

\lefthead{{\sc Planck}/LFI Data Streams: Statistics and Compression}
\righthead{M. Maris, D. Maino, C. Burigana, F. Pasian}

\begin{document}

\title{
\vspace{-2cm}{{\hspace{10.cm}\normalsize
                 \begin{tabular}{r}
                         May 3, 2000 \\
                         OAT Int. Rep. 71/00 \\
                         OAT Pub. Num. 2140 \\
                         Submitted to PASP \\
                 \end{tabular}}}\\\vspace{1cm}
 Data Streams from the Low Frequency Instrument\\
 On-Board the Planck Satellite:\\
 Statistical Analysis and Compression Efficiency
 }

\author{Michele Maris\altaffilmark{1}, Davide Maino\altaffilmark{1}, Carlo Burigana\altaffilmark{2}}
\and
\author{Fabio Pasian\altaffilmark{1} }

\altaffiltext{1}{Osservatorio Astronomico di Trieste, Via G.~B.~
Tiepolo 11, I-34131, Trieste, Italia. E-mail {\tt
<name>@ts.astro.it}}

\altaffiltext{2}{Istituto TeSRE, Consiglio Nazionale delle
Ricerche, Via Gobetti 101, I-40129, Bologna, Italia. E-mail {\tt
burigana@tesre.bo.cnr.it}
}


\received{by PASP May 3, 2000}
\revised{}
\accepted{}
\cpright{type}{year}
\journalid{VOL}{JOURNAL DATE}
\articleid{START PAGE}{END PAGE}
\paperid{MANUSCRIPT ID}
\ccc{CODE}


\authoremail{maris@ts.astro.it}

\begin{abstract}
The expected data rate produced by the Low Frequency Instrument
(LFI) planned to fly on the ESA Planck mission in 2007, is over a
factor 8 larger than the bandwidth allowed by the spacecraft
transmission system to download the LFI data. We discuss the
application of lossless compression to Planck/LFI data streams in
order to reduce the overall data flow. We perform both theoretical
analysis and experimental tests using realistically simulated data
streams in order to fix the statistical properties of the signal
and the maximal compression rate allowed by several lossless compression
algorithms.
We studied the influence of signal
composition and of acquisition parameters on the compression rate
\Cr\ and develop a semiempirical formalism to account for it.
The best performing compressor tested up to now is the arithmetic
compression of order 1, designed for optimizing the compression of
white noise like signals,
which allows an overall compression rate
$\overline{\Cr} = 2.65 \pm 0.02$.
We find that such result is not improved by other lossless compressors,
being the signal almost white noise dominated.
Lossless compression algorithms alone will not solve
the bandwidth problem but needs to be combined with other
techniques.
\end{abstract}


\section{Introduction and Scanning Strategy}
\label{sec:introduction}

The {\sc Planck} satellite (formerly COBRAS/SAMBA,
\cite{Bersanelli:etal:1996}), which is planned to be launched in
2007, will produce full sky CMB maps with high accuracy and
resolution over a wide range of frequencies
(\cite{Mandolesi:etal:1998a,Puget:etal:1998}).
Table~\ref{tab:PlanckTable}\ summarizes the basic properties of LFI
aboard Planck. The reported sensitivities per resolution element
-- i.e. a squared pixel with side equal to the Full Width at Half
Maximum (FWHM) extent of the beam --, in terms of antenna
temperature, represents the goals of LFI for 14 months of routine
scientific operations) as recently revised by the LFI Consortium
(\cite{Mandolesi:etal:1999}).

The limited bandwidth reserved to the downlink of scientific data
calls for huge lossless compression, theoretical upper limit being
about four (\cite{Maris:etal:SAIT}). Careful simulations are demanded to
quantify the capability of true compressors for ``realistic''
synthetic data and improve the theoretical analysis, including CMB
signal (monopole, dipole and anisotropies), foregrounds and
instrumental noise.

During the data acquisition phase the Planck satellite will rotate
at a rate of one circle per minute around a given spin axis that
changes its direction every hour (of 2.5$'$ on the ecliptic plane
in the case of simple scanning strategy), thus observing the same
circle on the sky for 60 consecutive times
(\cite{Mandolesi:etal:1998a,Mandolesi:etal:1998b}). LFI
will produce continuous data streams of temperature
differences between the microwave sky and a set of on-board
reference sources; both differential measurements and reference
source temperatures must be recorded.

The LFI Proposal assumes a sampling time $\tau_{\mathrm{s}} \sim
7$~msec for each detector (\cite{Mandolesi:etal:1998a}), thus
calling for a typical data rate of $\sim 260$~Kb/sec, while the
allocated bandwidth to download Planck data to ground is in total
$\sim 60$~Kb/sec. Assuming the total bandwidth to be equally split
between instruments, $\approx 30$\ Kb/sec on the average would be
assigned to LFI asking for a compression of about a factor $8.4$.
Data have to be downloaded without information losses and by
minimizing scientific processing on board.

A possible solution would be to adapt the sampling rate to the
angular resolution specific for each frequency. This should allow
to save about up to a factor $\approx 9$\ for the 30 GHz channel,
but since only $\approx 7\%$\ of the samples come from such
channel (see table \ref{tab:PlanckTable}) the overall reduction in
the final data rate would be $\approx 17\%$.

On the other hand, it is unlikely that the bandwidth for the
downlink channel may be enhanced to solve the bandwidth problem,
since the ground facilities are shared between different missions
and there is the need to minimize possible cross-talks between the
instrument and the communication system.

With the aim of optimizing of the transmission
bandwidth dedicated to the downlink of LFI data from the Planck
spacecraft to the FIRST/Planck Ground Segment, we
analyze in detail the role that can be played by
lossless compression of LFI data before they are sent to Earth.

We apply different compression algorithms to suitable sets of
Planck LFI simulated data streams generated by considering
different combinations of astrophysical and instrumental signals
and for different instrumental characteristics and detection
electronics.

The first considered contribution is that introduced by receiver
noise: we consider here the case of pure white noise and of white
noise coupled to $1/f$  noise with different knee frequencies.
The reference load temperature is assumed
to be 20~K for present tests; because of the strong
dependence of the $1/f$ noise on the load temperature,
this can be considered a worst case,
since the actual baseline reference load is of 4~K.

Different sky signal sources are subsequently added to the
receiver noise: CMB fluctuations, CMB dipole, Galaxy emission and
extragalactic point sources.
The signal from the different sky components are convolved
with the corresponding antenna pattern shapes, assumed to be
symmetric and gaussian with the FWHM reported in Table~1.

We generate simulated
data streams at the two extreme frequency channels, 30 GHz and 100
GHz and consider data streams with
different time lengths.

Regarding the detection electronics, we explore different signal
offset and scaling.

\noindent
The large number of above combinations
was systematically explored using an automated program generator
as described by \cite{Maris:etal:1998}.

\noindent
  In Section \ref{sec:components} we characterize quantitatively
the LFI signal component by component.
    Section \ref{sec:acquisition} we discuss how the acquisition chain is modeled to perform
compression simulations.
 A theoretical analysis of the compression
efficiency is presented in section \ref{sec:theorethical}.
 While section \ref{sec:statistical} is devoted to the analysis of
the signal statistics.
 The subject of quantization error is illustrated in section
\ref{sec:quantum}.
 The experimental protocol and results about compression are reported
in section \ref{sec:evaluation:and:results}.
 Further constraints on the on-board data compression are reported
in section \ref{sec:packeting}.
 A proposal for an alternative coding method is made in section \ref{sec:coding:scheme}.
 The overall compression rate is estimated in section \ref{sec:overall:cr}.
  Conclusions are in section \ref{sec:conclusions}.
  Appendix  \ref{appendix:a} is
included to further illustrate the estimation of the overall compression
rate.

\section{Characterization of Planck/LFI signal
components}\label{sec:components}

The simulated cosmological and astrophysical components are
generated according to the methods
described in \cite{Burigana:etal:1998b} and the data stream
and noise generation as in \cite{Burigana:etal:1997b},
\cite{Seiffert:etal:1997} and \cite{Maino:etal:1999}.
We summarize here below the basic points.

$\bullet$ {\it Modeling the CMB pattern} -- The CMB monopole and
dipole have been generated by using the Lorentz invariance of
photon distribution functions, $\eta$, in the phase space
(Compton--Getting effect): $\eta _{obs} (\nu_{obs},\vec n) = \eta
_{CMB} (\nu _{CMB}) \,$, where $\nu _{obs}$ is the observation
frequency, $\nu _{CMB} = \nu _{obs} (1+\vec \beta \times \vec n) /
\sqrt{1-\beta^2}$ is the corresponding frequency in the CMB rest
frame, $\vec n$ is the unit vector of the photon propagation
direction and $\vec \beta = \vec v/c$ the observer velocity. A
blackbody spectrum at $T_0=2.725$~K (\cite{Mather:etal:1999}) is
assumed for $\eta$. For gaussian models, the CMB anisotropies at
$l \ge 2$ can be simulated by following the standard spherical
harmonic expansion (see, e.g., \cite{Burigana:etal:1998} or by using
FFT (Fast Fourier Transform) techniques which take advantage of
equatorial pixelisations
(\cite{Muciaccia:etal:1997})).

$\bullet$ {\it Modeling the Galaxy emission} -- The Haslam map at
408~MHz (\cite{Haslam:etal:1982}) is the only full-sky map
currently available albeit large sky areas are sampled at 1420~MHz
(\cite{Reich:1986}) and at 2300~MHz (\cite{Jonas:etal:1998}). To
clean these maps from free-free emission we use a 2.7~GHz
compilation of $\sim$ 7000 HII sources
(\cite{witebsky:etal:1978}), private communication) at resolution
of $\sim$~1$^{\circ}$. They are subtracted for modelling the
diffuse components and then re-added to the final maps. We use a
spectral index $\beta_{ff} = 2.1$ from 2.7 to 1~GHz and
$\beta_{ff} = 0$ below 1~GHz. We then combine the synchrotron maps
producing a spectral index map between 408-2300~MHz with a
resolution of $\lsim 2^{\circ} \div 3^{\circ}$ ($<\beta_{sync}>
\sim 2.8$). This spectral index map is used to scale the
synchrotron  component down to $\sim$ 10~GHz. In fact,  for
typical (local) values of the galactic magnetic field ($\sim
2.5\mu$G), the knee in the electron energy spectrum in cosmic rays
($\sim$ 15 Gev) corresponds to $\sim$~10~GHz
(\cite{Platania:etal:1998}). From the synchrotron map obtained at
10~GHz and the DMR 31.5~GHz map we derive a high frequency
spectral index map for scaling the synchrotron  component up to
Planck frequencies. These maps have a poor resolution and the
synchrotron structure needs to be extrapolated to Planck angular
scales. An estimate of the synchrotron angular power spectrum and
of its spectral index, $\gamma$ ($C_l \propto l^{-\gamma}$), has
been provided by \cite{Lasenby:etal:1998}; we used $\gamma = 3$
for the angular structure extrapolation
(\cite{Burigana:etal:1998}). Schlegel (\cite{schlegel:etal:1998})
provided a map of dust emission at 100$\mu$m merging the DIRBE and
IRAS results to produce a map with IRAS resolution ($\simeq 7'$)
but with DIRBE calibration quality. They also provided a map of
dust temperature, $T_d$, by adopting a modified blackbody
emissivity law, $I_\nu \propto B_\nu(T_d) \nu^{\alpha}$, with
$\alpha =2$. This can be used to scale the dust emission map to
Planck frequencies using the dust temperature map as input for the
$B_\nu(T_d)$ function. Unfortunately the dust temperature map has
a resolution of $\simeq 1^{\circ}$; again, we use an angular power
spectrum $C_l \propto l^{-3}$ to scale the dust skies to the
Planck proper resolution. Merging maps at different frequencies
with different instrumental features and potential systematics may
introduce some internal inconsistencies. More data on diffuse
galactic emission, particularly at low frequency, would be
extremely important.

$\bullet$ {\it Modeling the extragalactic source fluctuations} --
The simulated maps of point sources have been created by an
all--sky Poisson distribution of the known populations of
extragalactic sources in the $10^{-5}< S(\nu)< 10$ Jy flux range
exploiting the number counts of \cite{Toffolatti:etal:1998} and
neglecting the effect of clustering of sources. The number counts
have been calculated by adopting the \cite{Danese:etal:1987}
evolution model of radio selected sources and an average spectral
index $\alpha=0$ for compact sources up to $\simeq 200$ GHz and a
break to $\alpha=0.7$ at higher frequencies (see \cite{Impey:1988};
\cite{DeZotti:1998}), and by the model C
of \cite{Franceschini:etal:1994} updated as in
\cite{Burigana:etal:1997a}, to account for the isotropic sub-mm component
estimated by \cite{Puget:etal:1996} and \cite{Fixsen:etal:1996}. At
bright fluxes, far--IR selected sources should dominate the number
counts at High Frequency Instrument (HFI) channels for $\nu \gsim
300$~GHz, whereas radio selected sources should dominate at lower
frequencies (\cite{Toffolatti:etal:1998}).

$\bullet$ {\it Instrumental noise} -- The white noise depends on
instrumental performances (bandwidth $\Delta \nu$, system temperature
$T_{sys}$), on
the observed sky signal, $T_{sky}$, dominated by CBM monopole, and on the
considered integration time, $\tau$, according to:

\begin{equation}
\Delta T_{\rm wn} = \frac{\sqrt{2}(T_{sys} + T_{sky})}{\sqrt{\Delta \nu \, \tau}} \, .
\label{whitenoise}
\end{equation}

Under certain idealistic assumptions, \cite{Burigana:etal:1997b}
and \cite{Seiffert:etal:1997} provide analytical estimates for the
knee frequency, $f_k$, of LFI radiometers; it is predicted to
critically depend also on the load temperature, $T_{load}$, according to:

\begin{equation}
f_k = \frac{A^2{\Delta \nu}}{8}(1-r)^2\left(\frac{T_{sys}}{T_{sys}+T_{sky}}\right)^2 \, ,
\label{knee}
\end{equation}

\noindent
where $r=(T_{sky}+T_{sys})/(T_{load}+T_{sys})$ and $A$ is a constant, depending
on the state of art of radiometer technology, which
has to be minimized for reducing via hardware the knee frequency
(current estimates are $A\sim 1.8\times 10^{-5}$ for 30 and 44 GHz radiometers and
$A\sim 2.5\times 10^{-5}$ for 70 and 100 GHz).

Recent experimental results from Seiffert (private
communication \cite{seiffert:99}) show knee frequency values of this order of
magnitude, confirming that the present state of art of the
radiometer technology is close to reach the ideal case.

A pure white noise stream can be easily generated by employed
well tested random generator codes and normalizing their output to the white
noise level $\Delta T_{\rm wn}$.
A noise stream which takes into account both white noise and $1/f$ noise
can be generated by using FFT methods.
After generating a realisation of
the real and imaginary part of the Fourier coefficients with
spectrum defined $S_{noise}(f) \propto (1 + f_k/f)$,
we transform
them and obtain a real noise stream which has to be normalized to the white
noise level $\Delta T_{\rm wn}$
(\cite{Maino:etal:1999}).

$\bullet$ {\it Modeling the observed signal} --
We produce full sky maps, $T_{sky}$, by adding the antenna temperatures from
CMB, Galaxy emission and extragalactic source fluctuations.
Planck will
perform differential measurements and not absolute temperature
observations; we then represent the final observation in a given
$i$-$th$ data sample in the form $T_i = R_i (T_{sky,i} + N_i -
T_{x,i}^r) \,$, where $N_i$ is the instrumental noise generated
as described above.

$T_{x,i}^r$ is a reference temperature
subtracted in the differential data and $R_i$ is a constant which
accounts for the calibration. Of course, the uncertainty on $R_i$
and the non reduced time variation of $T_{x,i}^r$ have to be much
smaller than the Planck nominal sensitivity. Thus, we generate the
``observed'' map assuming a constant value, $T_x^r$, of
$T_{x,i}^r$ for all the data samples. We note that possible constant
small off-sets in $T_{x}^r$ could be in principle accepted, not
compromising an accurate knowledge of the anisotropy pattern.
We arbitrarily generate the
``observed'' map with $R_i=R=1$ for all the data samples.

\section{A model of Acquisition Chain}\label{sec:acquisition}

To test rigorously the efficiency of different compressors the
best solution is to generate a realistically simulated
signal for different mission hypotheses and apply to them the
given compressors. To be realistical the simulation of the
signal generation should contain both astrophysical and
instrumental effects. It would be helpful that the final
simulation would be able to given a hint about the influence of
the various signal components and their variance.
 Of course it is useless to reproduce in full detail the LFI
to obtain a signal simulation accurate enough to test
compressors. A simplified model of the LFI, its
front-end electronics and its operations will be enough.

At the base of the simplified model is the concept of {\em
acquisition pipeline}. This pipeline is composed by all the
modules which process the astrophysical signal: from its
collection to the production of the final data streams which are
compressed and then sent to Earth. In the real LFI, the
equivalent of the acquisition pipeline may be obtained following
the flow of the astrophysical information,
from the telescope through the front-end electronics
and the main Signal Processing Unit (SPU) to the memory of the
Data Processing Unit (DPU) which is in charge to downlink it to
the computer of the spacecraft and then to Earth. The acquisition
pipeline is represented in figure \ref{fig:acquisition:pipe}.
Since its purpose is to describe the signal processing and its
parameters it must not be regarded as a representation of the true
on-board electronics since some functionalities may be shared
between different real modules.
 In this scheme {\em Front End} operations of the true LFI are
assigned to the first simulation level, while on-board processing
and compression to the second one.

The simulated microwave signal from the sky is collected and
compared with the temperature of a reference load which, in our
simulations, is supposed to have exactly the CMB temperature
$T_0=2.725$~K (\cite{Mather:etal:1999})
\footnote{Alternatively, sky
the reference-load signals may be sampled separately
and then $\Delta T$\ may be compute numericaly by the DPU.}.
The difference $\Delta T$\ expressed in $\mu$K\ is sampled along a
scan circle producing a data stream of 60 scan circles with 8640
samples (pointings).

Signal detection is simulated by
\cite{Bersanelli:etal:1996,Maris:etal:ADASS,Maris:etal:SAIT}

\begin{equation}\label{eq:acquisition}
 \Vout = \AFO + \VOT \cdot \Delta T,
\end{equation}

\noindent where \Vout\ is the detection chain output in Volts,
\VOT\ is the antenna temperature to the detector voltage
conversion factor ($-0.5$V/K $\leq VOT \leq +1.5$ V/K) while \AFO\
is a detection chain offset ($-5$V $\leq \AFO \leq +5$V). Of
course in our simulation this offset takes into account all offset
sources, including variations of the reference temperature, and
not only of the electrical offset. Similarly the \VOT\ factor
takes into account also differences among the different
detectors which affect the calibration of the temperature/voltage
relation. The range for \VOT\ and \AFO\ is large enough to
include the whole set of nominal instrumental configurations,
allowing also for somewhat larger and smaller values.

The analog to digital conversion (ADC) is described by the formula:

\begin{equation}\label{eq:digitizzation}
   \Voutadu \mathrm{(adu)} =
        \trunc
        \left( 2^{\Nbits} \; \cdot \; \frac{\Vout - \Vmin}{\Vmax - \Vmin} \right),
\end{equation}

\noindent where $\trunc(.)$\ is the decimal truncation operator,
\Nbits\ is the number of quantization bits produced by the
ADC, while $\Vmin$\ and $\Vmax$\ are the
lower and upper limits of the voltage scale accepted in input by
the ADC. In our case: $\Nbits = 16$\ bits, $\Vmin = -10$\ V, $\Vmax = +10$\
V. So the quantization unit ``adu'' (analog/digital unit) is

\begin{equation}\label{eq:adu}
1 \, \adu = \frac{\Vmax - \Vmin }{2^{\Nbits} }
\end{equation}

\noindent or in terms of antenna temperature the quantization step is

\begin{equation}\label{eq:adu:K}
 \Delta = \frac{\Vmax - \Vmin }{2^{\Nbits} \VOT}
\end{equation}

\noindent for a typical $\VOT = 1$~V/K, $\Nbits = 16$\
bits, $1\; \Delta \approx 3 \times 10^{-4}$~K/adu.
After digitization the simulated signal is written into a binary
file of 16 bits integers and sent to the compression pipeline.

The simplified LFI is composed of four acquisition
pipelines, one for each frequency, each one being representative
of the set of devices which form the full detection channel for
the given frequency. The overall data-rate after loss-less compression
for LFI should be obtained summing the contribution expected from each
detector. Since in the real device each radiometer for a
given frequency channel, will be characterized by different values
of \VOT\ and \AFO, the distribution of these parameters has to be
taken in account computing the overall compression efficiency. In
particular a greater attention should be devoted to the
distribution of the \VOT\ parameter since the
compression efficiency is particularly sensitive to it. However,
since the distribution of operating conditions and instrumental
parameters are not yet fully defined, we assumed that all the
detectors belongin to a given frequency channel are identical
\footnote{But see section \ref{sec:overall:cr} and the
appendix for a more detailed discussion.} and located at the telescope focus.

\section{An Informal Theoretical Analysis About the Compression
Efficiency}\label{sec:theorethical}

An informal theoretical analysis may be helpful to evaluate the maximum
lossless compression efficiency expected from LFI
and to discuss the behaviour of the different compressors.
For further details we remind the reader to \cite{Compression:Book}.

Data compression is based on the partition of a stream of
bits into short chunks, represented by strings of bits of
fixed length \LIn, and to code each string of bits \stringIn\ into
another string \stringOut\ whose length \LOut\ is variable and, in
principle, shorter than \stringIn. In this scheme, when the string
of bits represents a message, the possible combinations of bits in
\stringIn\ represents the {\em symbols} by which the {\em message}
is encoded. From this description the compression operation is
equivalent to map the input string set \SetstringIn\ into an
output string set \SetstringOut\ through a {\em compressing
function} \cmpfunc . A compression algorithm is called {\em
lossless}\ when it is possible to reverse the compression process
reconstructing the \stringIn\ string from \stringOut\ through a
{\em decompression} algorithm. So the condition for a compression
programs to be lossless is that the related \cmpfunc\ is a
one-to-one application of \SetstringIn\ into \SetstringOut. In
this case the {\em decompressing algorithm}\ is the inverse
function of \cmpfunc. Of course in the general case it is not
possible to have at the same time lossless compression and $\LIn >
\LOut$\ for any string in the input set. The problem is solved
assuming that the discrete distribution $P(\stringIn)$\ of strings
belonging to the input stream of bits is not flat but that a most
probable string exists. So a good \cmpfunc\ will assign the
shortest \stringOut\ to the most probable \stringIn\ and, the
least probable the input string, the longest the output string. In
the worst case output strings {\em longer} than the input string
will be assigned to those strings of \SetstringIn\ which are least
probable. With this statistical tuning of the compression function
the final length of the compressed stream will be shorter than the
original length, the averaged length of $\stringOut$\ being:

\begin{equation}
 \ALOut = \sum_{\stringIn \in \SetstringIn} P(\stringIn)
 \LOut(\cmpfunc(\stringIn)).
\end{equation}

Several factors affect the efficiency of a given compressor, in particular
 best performances are obtained when the compression algorithm is tuned
on the specific distribution of symbols. Since the symbol distribution
depends on \LIn\ and on the specific input stream,
an ideal general-purpose self-adapting compressor should be able
to perform the following operations: {\it i)} acquire the full bit
stream (in the hypothesis it has a finite length) and divide it in
chunks of length \LIn, {\it ii)} perform a frequency analysis of the
various symbols, {\it iii)} create an optimized {\em coding table} which
associates to each \stringIn\ a specific \stringOut,
{\it iv)} perform the compression according to the optimized coding
table, {\it v)} send the coding table to the uncompressing program
together with the compressed bit stream. The uncompressing program
will restore the original bit stream using the associated
optimized coding table.

In practice in most cases the chunks size \LIn\ is hardwired into
the compressing code (typically $\LIn = 8$\ or 16 bits), also the
fine tuning of the coding table for each specific bit stream is too
expensive in terms of computer resources to be performed in this
way, and the same holds for coding table transmission. So
there are compressors which work as if the coding table or,
equivalently, the compression function is fixed. In this way the
bit stream may be compressed chunk by chunk by the compressing
algorithm which will act as a filter. Other compressors perform the
statistical tuning on a small set of chunks taken at the beginning
of the stream, and then apply the same coding table to the full
input stream. In this case the compression efficiency will be
sensitive to the presence of correlations between difference parts
of the input stream. In this respect self-adaptive codes may be
more effective than non-adaptive ones, if their adapting strategy
is sensitive to the kind of correlations in the input
stream.

On the other hand other solutions may be adopted to obtain a good
compromise between computer resources and compression
optimization. For example all of the previous compressors are
called {\em static} since the coding table is fixed in one way or
the other at the beginning of the compression process and then
used all over the input stream. Another big class of self-adaptive
codes is represented by {\em dynamical} self-adaptive compressors,
which gain the statistical knowledge about the signal as the
compression proceeds changing time by time the coding table. Of
course these codes compress worse at the beginning and better at
the end of the data stream, provided its statistical properties
are stationary. They are also able to self-adapt to remarkable
changes in the characteristics of the input stream,
but only if these changes may be sensed by the
adapting code. Otherwise the compressor will
behave worse than a well-tuned static compressor. Moreover, if the
signal changes frequently, it may occur that the advantage of the
dynamical self adaptability is compensated by the number of
messages added to the output stream to inform the decompressing
algorithm of the changes occurred to the coding table. Last but
not least, if some error occurs during the transmission of the
compressed stream and the messages about changes in the coding
table are lost, it will be impossible to correctly restore it at
the receiving station. This problem may be less severe for a
static compressor since, as an example, it is possible to split
the output stream in packets putting {\em stop codes} and storing
the coding table on-board until a {\em confirmation message} from
the receiving station is sent back to confirm the correct
transmission.

It is then clear that each specific
compression algorithm is {\em statistically optimized}\ for a given
kind of input stream with its own statistical properties. So to
obtain an optimized compressor for LFI it is
important to properly characterize the statistics of the signal to
be compressed and to test different existing compressors in order
to map the behaviour of different compression schemes using
realistically simulated signals and, as soon as possible, the
true signals produced by the LFI electrical model.

In order to evaluate the performances of different compression
scheme we considered the {\em Compression Rate}\ \Cr\ defined as:

\begin{equation}\label{Cr:Definition}
\Cr = \frac{L_u}{L_c}
\end{equation}

\noindent where $L_u$\ is the length of the input string in bytes
and $L_c$\ is the length of the output string in bytes
\footnote{Often compressors are evaluated looking at the
compression efficiency $\eta_c = 1/C{\rm r}$\ but we considered \Cr\
more effective for our purposes.}. Other important estimators of
to evaluate the performances of a given compression code
are the memory allocation and the compression time. Both of them
must be evaluated working on the final model of the on board computer.
Since this component is not fully defined for the Planck/LFI mission,
in this work we neglect these aspects of the problem.

The measure represented by one of the 8640 samples which form one
scan circle is white noise dominated, the r.m.s.
$\sigma_T$\ being about a factor of ten higher then the CMB
fluctuations signal. If so, at the first approximation it is
possible to assume the digitized data stream from the front-end
electronics as a stationary time serie of independent samples
produced by a normal distributed white noise generator. In such
situation symbols are represented by the quantized signal levels,
and it is easy to infer the best coding table and by the
information theory the expected compression rate for an optimized
compressor is promptly estimated (\cite{Gaztnaga:etal:1998}). In our
notation, for a zero average signal:

\begin{equation}\label{eq:cr:theory}
 \CrTh = \frac{\Nbits \ln 2}{\ln(\sqrt{2\pi e} \sigma_l/\adu) +\ln \VOT}
\end{equation}

\noindent where $\sigma_l$\ is the r.m.s. of the sampled signal
\footnote{It has to be noted that eq. (\ref{eq:cr:theory}) is an
approximated formula which is rigorously valid when $\sigma_l/\adu
\gg 1$.}.

From Eq.~(\ref{eq:cr:theory}) it is possible to infer that the higher is
the \VOT, (i.e. higher is the $\Delta T$\ resolution) the worse is the
compression rate, as already observed in
\cite{Maris:etal:ADASS}, \cite{Maris:etal:SAIT}. The reason being the
fact that as \VOT\ is increased the number of quantization levels
(i.e. of symbols) to be coded is increased and their distribution becomes more
flat increasing \ALOut.
 Assuming that all the white noise is thermal in origin $\sigma_l
\approx \sigma_T \approx 2 \times 10^{-3}$\ K. With the \adu\
defined in equation (\ref{eq:adu}) together with the typical
values of \Vmin\ and \Vmax\ assumed therein and $\Nbits = 16$\
bits we have $\CrTh \sim 11.09/(3.30+\ln\VOT)$. In conclusion, for $\VOT =
0.5$, $1.0$, $1.5$\ V/K the \CrTh\ is respectively $4.26$, $3.36$,
$3.00$. In addition figure \ref{fig:CrTh}\ represents the effect
of a reduction of \Nbits\ on \CrTh\ compared to \CrTh\ for
$\Nbits=16$.

\section{Statistical Signal Analysis}\label{sec:statistical}

A realistic estimation of the compression efficiency must be based
on a quantitative analysis of the signal statistics, which
includes: statistics of the binary representation (section
\ref{sec:Binary}), entropy {section \ref{sec:Entropy}) and
normality tests (section \ref{sec:Normality}).

\subsection{Binary Statistics}\label{sec:Binary}
Most of the off-the-shelf compressors considered here do not
handle 16 bits words, but 8 bits words. The 16 bits
samples produced by the adc unit are splitted into two consecutive
8 bits (1 byte) words labeled: {\em most significant bits} (MSB)
word and {\em least significant bits} (LSB) word. To properly
understand the compression efficiency limits it is important to
understand the statistical distribution of 8 bits words composing
the quantized signal from LFI.

Figure \ref{fig:qstat}\ represents the frequency distribution of
symbols when the full data stream of 60 scan circles is divided
into 8 bits words. Since for most of the samples the range spans
over $\approx 64$\ levels (5 bits) only the bytes corresponding to
the MSB words assume a limited range of values producing the
narrow spike in the figure. The belt shaped distribution at the
edges is due to the set of LSB words.
 The distributions are quite sensitive to the quantization step,
but do not change too much with the signal composition, the largest
differences coming from the cosmological dipole contribution.

From the distribution in figure \ref{fig:qstat}\ one may wonder if
it would not be possible to obtain a more effective compression
splitting the data stream into two substreams: the MSB substream
(with compression efficiency \CrMSB) and the LSB substream (with
compression efficiency \CrLSB). Since the two components are so
different in their statistics, with the MSB substream having an
higher level of redundancy than the original data stream, it would
be reasonable to expect that the final compression rate
$2/(1/\CrMSB + 1/\CrLSB)$\
be greater than the compression rate obtained compressing
directly the original data stream. We tested this procedure taking
some of the compressors considered for the final test. From these
tests It is clear that $\CrMSB >> \Cr$\ but since most of the
redundancy of the original data stream is contained in the MSB
substream the LSB substream can not be compressed in an effective
way, as a result $\CrLSB < \Cr$\ and $2/(1/\CrMSB + 1/\CrLSB) \lsim \Cr$.
So the best way to perform an efficient compression is to apply
the compressor to the full stream without performing the MSB / LSB
separation. Apart from these theoretical considerations, we
performed some tests with our simulated data stream confirming
these result.

\subsection{Entropy Analysis}\label{sec:Entropy}

Equation (\ref{eq:cr:theory}) is valid in the limit of a
continuous distribution of quantization levels. Since in our case
the quantization step is about one tenth of the signal rms this is
no longer true. To properly estimate the maximum compression
rate attainable from these data we evaluate the entropy of the
discretized signal using different values of the \VOT.

Our entropy evaluation code takes the input data stream and
determines the frequency $f_s$\ of each symbol $s$\ in the
quantized data stream and computing the entropy as: $- \sum_s f_s
\log_2 f_s$\ where $s$ is the symbol index. In our simulation we
take both 8 and 16 bits symbols ($s$\ spanning over $0$, $\dots$,
$255$\ and $0$, $\dots$, $65535$). Since in our scheme the ADC
output is 16 bits, we considered 8 bits symbols entropy both for
the LSB and MSB 8 bits word and 8 bits entropy after merging the
LSB and MSB significant bits set.

As expected, since \AFO\ merely shifts the quantized signal
distribution, entropy does not depend on \AFO. For this reason we
take $\AFO = 0$ V, i.e., no shift.

Table \ref{tab:entropy}\ reports the 16 bits entropy as a function
of \VOT, composition and frequency. As obvious entropy, i.e.
information content, increases increasing \VOT\ i.e. quantization
resolution. The entropy $H$ distribution allows to evaluate the \Cr\
r.m.s. espected from different data streams realizations:

\begin{equation}\label{eq:cr:rms}
 \mathrm{RMS}(\Cr) \approx \Cr \frac{\mathrm{RMS}(H)}{H}.
\end{equation}

Since data will be packed in chuncks of finite length it is important
not only to study the entropy distribution for the entire data-stream,
which will give an indication of the overall compressibility of the
data stream as a wall, but also the entropy distribution for short
packets of fixed length.
So each data stream was splitted into an integer number of chunks of
fixed length $l_{\mathrm{chunck}}$. For each chunck the entropy
was measured, and the corresponding distribution of entropies for the
given \Lchunck\ as its mean and rms was obtained. We take
$l_{\mathrm{chunk}} = 16$, $32$, $64$, $135$, $8640$, $17280$
16-bits  samples, so each simulated $8640 \times 60$\ data stream
will be splitted into 32400, 16200, 8100, 3840, 60, 30 chuncks.
Small chunck sizes are introduced to study the entropy distribution
as seen by most of the true compressors which do not compress one circle
(8640 samples) at a time. Long chuncks distributions are usefull to
understand the entropy distribution for the overall data-stream.

The entropy distribution per chunck is approximately described by a
normal distribution (see figure \ref{fig:entropy:distribution}),
so the mean entropy and its r.m.s. are enough to characterize the results.
Not however that the corresponding distribution of compression rates
is not exactly normally distributed, however for the sake of this analysis
we will assume that even the \Cr\ distribution is normally distributed.

The mean entropy measured over one scan circle
($l_{\mathrm{chunk}} = 8640$\ samples) coincides with the entropy
measured for the full set of 60 scan circles, the entropy r.m.s.
being of the order of $10^{-2}$\ bits. Consequently the expected
r.m.s. for \Cr\ compressing one or more circles at a time will be
less than $1\%$.

The mean entropy and its rms are not independent quantities.
Averaged entropy decreases as \Lchunck\ decreases, but
correspondingly the entropy r.m.s. increases. As a consequence
the averaged \Cr\ decreases decreasing \Lchunck, but the fraction
of chunks in which the compressor performs significantly worst than
in average increases. The overall compression rate,
i.e. the \Cr\ referred to the full mission, beeing affected by them.

\subsection{Normality Tests}\label{sec:Normality}

Since normal distribution of signals is assumed in \ref{sec:theorethical}\
it would be interesting to fix how much the digitized signal distribution
deviates  from the normality.
Also it would be important to characterize the influence
of the 1/f noise and of the other signal components, especially
the cosmic dipole, in the genesis of such deviations. To obtain an
efficient compression it would be important that the samples are
as more as possible statistically uncorrelated and normally
distributed. In addition one should make sure that the detection
chain does not cause any systematic effect which will introduce
spurious non normal distributed components. This is relevant not
only for the compression problem itself, which is among the data
processing operations the least sensitive to small deviations from
the normal distribution, but also in view of the future data
reduction, calibration and analysis. For them the hypothesis of
normality in the signal distribution is very important in order to
allow a good separation of the foreground components. Last but not
least, the hypothesis of conservation of normality along the
detection chain, is important for the scientific interpretation of
the results, since the accuracy expected from the Planck/LFI
experiment should allow to verify if really the distribution of
the CMB fluctuations at $l \gsim 14$\ is normal, as predicted by
the standard inflationary models, or as seems suggested by
recent 4 years COBE/DMR results
(\cite{Bromley:Tegmark:1999,Ferreira:etal:1999}).

For this reason a set of normality tests was applied to the
different components of the simulated signal before and after
digitization in order to characterize the signal statistics and
its variation along the detection process. Of course this work may
be regarded as a first step in this direction, a true calibration
of the signal statistics will be possible only when the front end
electronics simulator will be available. Those tests have
furthermore the value of a preparation to the study of the true
signal.

Normality tests were applied on the same data streams used for
data compression. Given on board memory limits, it is unlikely
that more than a few circles at a time can be stored before
compression, so statistical tests where performed regarding each
data stream for a given pointing, as a collection of 60
independent realizations of the same process. Of course this is
only approximately true. The 1/f noise correlates subsequent scan
circles, but since its r.m.s. amplitude per sample is typically
about one-tenth of the white noise r.m.s. or less, these
correlations can be neglected in this analysis.

Starting from the folded data streams a given normality test was
applied to each set of 60 realizations for each one of the 8640
samples, transforming the {\em stream of samples} in a {\em stream
of test results} for the given test. The cumulative distribution
of frequency was then computed over the 8640 test results. Since
60 samples does not represent a large statistics, significant
deviations from theorethically evaluated confidence levels
are expected resulting in an excessive rejection or acceptation rates.
For this reason each test was calibrated applying it
to the undigitized white noise data stream.
Moreover, in order to analyze how the normality evolves increasing the
signal complexity, tests was repeated increasing the information content
of the generated data stream.

To simplify the discussion we considered as a reference test the
usual Kolmogorow - Smirnov D test from
\cite{Numerical:Recipes:1986}\ and we fix a $95\%$\ acceptance
level. The test was ``calibrated'' using the MonteCarlo white
noise generator of our mission simulator in order to fix the
threshold level $D_{\mathrm{th}}$\ as the $D$\ value for which
more than $95\%$\ of our samples show $D \leq D_{\mathrm{th}}$.
From Table \ref{tab:dtest}\ the quantization effect is evident, at
twice the nominal quantization step ($\VOT = 2$\ V/K)\ in $30\%$ of the
samples (i.e. 2592 samples) the distribution of realizations
deviates from a normal distribution ($D > D_{\mathrm{th}}$). Since
the theoretical compression rate from eq. (\ref{eq:cr:theory}) is
for a continuous distribution of levels ($\sigma \gg \Delta$) a
smaller \Cr\ should is expected. Since the deviation from the
normal distribution is a systematic effect, for the sake of
cosmological data analysis one may tune the D test to take account
of the quantization. As an example, the third line in Tab.
\ref{tab:dtest}\ reports the threshold for the quantized signal
$D_{\mathrm{th}}^{\mathrm{Q}}$\ for which $95\%$\ of the quantized
white noise samples are accepted as normal distributed. The line
below represents the success rate for the full quantized
signal. After the recalibration the test is able to recognize that
in $95\%$\ of the cases the signal is drawn from a normal
distribution, but at the cost of a growth in the threshold $D$\
which now is a function of the quantization step $\Delta$.

As for the entropy distribution and the binary statistics, even in this case
most of the differences between the results obtained for a pure white
noise signal and the full signals are explained by the presence of the
cosmological dipole. However these
simulations are not accurate enough to draw any quantitative
conclusions about the distortion in the sampling statistics
induced by digitization, but they suggest that to
approximate the instrumental signal as a quantized white noise
plus a cosinusoidal term associated to the cosmic dipole
is more than adequate in order to understand
the optimal loss-less compression rate achievable in the
case of the Planck/LFI mission.

\section{Quantization and Quantization Error}\label{sec:quantum}

A possible solution to solve the bandwidth problem is to reduce
the amount of information of the sampled signal i.e. its entropy.
Independently from the way in which this is performed, the final
compression strategy will be lossy, and the final reconstructed
(uncompressed) signal will be corrupted with respect to the
original one, degrading in some regard the experimental
performances. In this regard, any sort of lossy compression may be
seen as a kind of signal rebinning with a coarser resolution
(quantization step) in $\Delta T/T$.

There are at least six aspects in Planck/LFI operations which may
be affected by a coarser quantization:

\begin{enumerate}
\item\label{item:DeltaCl}
$C_l$\ and periodical signals reconstruction;
\item
destriping;
\item
foreground separation;
\item
point like sources detection;
\item
variable sources characterization;
\item
tests for normality of CMB fluctuations.
\end{enumerate}

\noindent Since the non linear nature of the quantization process,
all of them are hard to be analytically evaluated and for this
reason a specific simulation task is in progress for the Planck/LFI
collaboration (\cite{White:Seyfert:1999}, \cite{Maris:etal:00}).
However an heuristic evaluation for the point
(\ref{item:DeltaCl}) by analytical means is feasible.

Quantization operates a convolution of the normal distribution of
the input signal with the quantization operator $(x:\Delta) =
\sign(x) \Delta*floor(|x/\Delta|)$. If the quantization error: $(x
- (x:\Delta))$\ is uniformly distributed its expectation is
$\Delta/2$\ and its variance is $\Delta/\sqrt{12}$\
(\cite{IEEE:PAPER}). Quantization over a large amount of samples
may be regarded as an extra source of noise which will enhance the
variance per sample. If the quantization error is statistically
independent from the input quantized signal and if it may be added
in quadrature to the white noise variance $\sigma_{WN}$, the total
variance per sample will be $\approx \sigma_{WN}^2 \left( 1 +
\frac{\Delta^2/\sigma_{WN}^2}{12}\right)$. So for $\Delta \lsim
\sigma_{WN}$\ the expected quantization r.m.s. is $\lsim 4\%$.
From error propagation the relative error on the $C_l$\ is
(\cite{Maino:PhD}):

\begin{equation}
    \frac{\delta C_l}{C_l} =
\sqrt{\frac{4 \pi}{A}}
\sqrt{\frac{2}{(2 l + 1)f_{sky}}}
           \left[ 1 + \frac{\sigma^2 \theta^2}{B_l^2 C_l} \right]
\end{equation}

\noindent so that the quantization contribution to the overall
error will be small and dominated by the cosmic variance for a
large set of $l$. However the application of such encouraging
result must be considered carefully in a true experimental
framework. Apart from the assumptions, it has to be demonstrated
indeed that a large quantization error like this will not harm
significantly the aforementioned aspects, moreover the impact of
signal quantization will depend on how and in which point of the
detection chain it will be performed.

\section{Experimental Evaluation of Off-The-Shelf
Compressors}\label{sec:evaluation:and:results}

This section describes the evaluation protocol and the
experimental results of the compression of simulated data streams
for Planck/LFI.

\subsection{Evaluation Protocol}\label{subsec:evaluation}

 First tests were performed on a HP-UX workstation on four compressors
(\cite{Maris:etal:ADASS})\ but given the limited number of
off-the-shelf compression codes for such platform, we migrated the
compression pipeline on a Pentium III based Windows/NT
workstation.

As described in section \ref{sec:components}\ the signal
composition is defined by many components, both astrophysical and
instrumental in origin. In particular, it is important to understand how
each component or instrumental parameter, introducing deviations
to the pure white noise statistics, affects the final compression
rate.

To scan systematically all the relevant combinations of signal
compositions and off-the-shelf compressors, a Compression Pipeline
was created. The pipeline is based on five main components: the
signal quantization pipeline, the signal database, the compression
pipeline, the compression data base, the post-processing pipeline.
The signal quantization pipeline performs the operations described
in the upper part of figure \ref{fig:acquisition:pipe}. The
simulated astrophysical signals are hold in a dedicated section of
the signal archive, they are processed by the quantization
pipeline and stored back in a reserved section of the signal
archive. So quantized data streams are generated for each relevant
combination of the quantization parameters, signal composition and
sky pointing.

Each compressor is then applied by the compression pipeline to the
full set of quantized signals in the signal archive. Results, in
terms of compression efficiency as a function of quantization
parameters are stored in the compression database. The statistical
analysis of section \ref{sec:statistical}\ are performed with a
similar pipeline.

Finally the post-processing pipeline scans the compression data
base in order to produce plots, statistics, tables and synthetic
fits. Its results are usually stored into one of the two
databases.

The pipeline is managed by {\tt PERL} 5.004 script files which drive
FORTRAN, C, IDL programs or on-the-shelf utilities gluing and
coordinating their activities. Up to $\approx 75\,000$\ lines of
repetitive code are required per simulation run. They are
generated by a specifically designed Automated Program Generator
(APG) written in IDL (\cite{Maris:etal:1998}).
 The APG takes as an input a table which
specifies: the set of compressors to be tested, the set of
quantization parameters to be used, the order in which to perform
the scan of each parameter/compressor, the list of supporting
programs to be used, other servicing parameters. The program
linearizes the resulting parameter space and generates the {\tt
PERL} simulation code or, alternatively, performs other operations such
as: to scan the results data base to produce statistics, plots,
tables, and so on. The advantage of this method is that a large
amount of repetitive code, may be quickly produced, maintained or
replaced with a minor effort each time a new object (compressor,
parameter or analysis method) is added to the system.

\subsection{Experimental Results}\label{subsec:results}

Purpose of these compression tests is to give an upper limit to
the lossless compression efficiency for LFI data and to look for
an optimal compressor to be proposed to the LFI consortium.

A decision about the final compression scheme for Planck/LFI has
not been taken yet and only future studies will be able to decide
if the best performing one will be compatible with on-board
operations (constrained by: packet independence and DPU
capabilities) and will be accepted by the Planck/LFI
collaboration.

For this reason up to now only off-the-shelf compressors and
hardware where considered. To test any reasonable compression
scheme a wide selection of lossless compression algorithms,
covering all the known methods, was applied to our simulated data.
Lacking a comprehensive criteria to fix a final compressor, as
memory and CPU constrains, we report in a compact form the results
related to all the tested compressors. We are confident that in
the near future long duration flight balloon experiments as
on-board electronics prototypes will provide us with a more solid
base to test and improve the final compression algorithms looking
at real data.

Tables
 \ref{tab:press:list:a},
 \ref{tab:press:list:c}
list the selected compression programs. Since the behaviour (and
efficiency) of each compressor is determined by a set of
parameters one or more {\em macro file}\ operating a given
combination of compressor code plus parameters is defined. It has
to be noted that {\tt uses}\ is a space qualified algorithm, based
on Rice compression method, for which space qualified dedicated
hardware already exists.

To evaluate the performances of each compressor, {\em figures of
merit}\ are drawn like the one in figure \ref{fig:merith}\ which
shows the results for the best performing compressor: {\tt
arith-n1}. Looking at such figures it is possible to note as the
compression efficiency does not depend much on the signal
composition. This is true even when large, impulsive signals, as
planets, affecting few samples over thousands are introduced.
Again, this is a consequence of the fact that white noise
dominates the signal, being the most important component to affect
the compression efficiency. In this regard it has been speculated
that the 1/f component should improve the correlation between
neighborhood samples affecting the compression efficiency
(\cite{Maris:etal:ADASS}) no relevant effect may be detected into our
simulations. As an example from figure \ref{fig:merith}\ for the
30 GHz signal the addition of the 1/f noise to the white noise
data stream affects the final \Cr\ for less than $0.5\%$.

The only noticeable (i.e. some $6\%$) effect due to an increase in
the signal complexity, occurs when the cosmic dipole is added. In
the present signal the dipole amplitude is comparable with the
white noise amplitude ($\approx 3$\ mK) so its effect is to
distort the sample distribution, making it leptocurtic. As a
consequence compressors, which usually work best for a normal
distributed signal, becomes less effective. Since the dipole
introduces correlations over one full scan circle, i.e. some
$10^3$\ samples, while compressors establish the proper coding
table observing the data stream for a small set of consecutive
samples (from some tens to some hundred samples), even a self
adaptive compressor will likely loose the correlation introduced
by the dipole. A proper solution to this problem is suggested in
section \ref{sec:coding:scheme}. The other signal components do
not introduce any noticeable systematic effect. The small
differences shown by the figures of merit may be due to the
compression variance and depend strongly on the compressor of
choice. As an example a given compressor may be more effective to
compress the simulated data stream with the full signal than the
associated simpler data stream containing only white noise, 1/f
noise, CMB and dipole. At the same time another compressor may
show an opposite behaviour.

As shown by Figure~\ref{fig:cr:ring}, and as expected from eq.
(\ref{eq:cr:theory}) increasing \VOT, i.e. increasing the
quantization step, increases the compression rate. In addition
\Cr\ increases increasing \Nc\ up to an $\approx 20 \%$. The
increase is noticeable for $\Nc < 15$\ and saturates after $\Nc =
30$. On the contrary its dependence on the offset (\AFO) is negligible
(less than $1\%$). For these reasons in the subsequent analysis
the AFO dependency is neglected and the corresponding simulations
are averaged.

\subsection{Synthetical Description}\label{sec:fit}
 The full data base of simulated compression results takes about 14
MBytes, for practical purposes it is possible to synthesize all
this information using a phenomenological relation which connects
\Cr\ with \Nc\ and \VOT\ whose free parameters may be fitted using
the data obtained from the simulations. In short:

\begin{equation}\label{eq:cr:fit}
\CrFit(\VOT, \Nc) =
    \frac{\CrOne}{{\mathcal{I}}(\Nc) +
    {\mathcal{S}}(\Nc) \ln \left[
    \frac{\mathrm{VOT}}{1\;\mathrm{V/K}}\right]}
\end{equation}

\noindent where \CrOne\ is the \Cr\ for $\Nc = 1$, $\VOT = 1.0$\
V/K, while ${\mathcal{I}(\Nc)}$\ and $\mathcal{S}(\Nc)$\ describe
the \Cr\ dependence on \Nc. In particular the relation is
calibrated for any compressor imposing that $\Cr(\VOT=1
\;\mathrm{V/K}, \Nc=1) = \CrOne$.

The linear dependency of $1/\CrFit$\ over $\ln \VOT$\ is a direct
consequence of equation (\ref{eq:cr:theory}), and is confirmed by
a set of tests performed over the full set of our numerical
results for the compression efficiency, the r.m.s. residual
between the best fit (\ref{eq:cr:fit}) and simulated data being
less than $1.5\%$, in almost the $92\%$\ of the cases and less
than $1\%$\ in $72\%$\ of the cases. The dependencies of its
parameters $\mathcal{I}$\ and $\mathrm{S}$\ over \Nc\ are obtained
by a test-and-error method performed on our data set and we did
not investigate further on their nature. For all practical
purposes our analysis shows that these functions are well
approximated by a series expansion:

\begin{equation}\label{eq:cr:fit:intercept}
{\mathcal{I}}(\Nc) \approx \exp \left( \sum_{k=1}^{2} A_k (\ln
\Nc)^k \right),
\end{equation}

\begin{equation}\label{eq:cr:fit:slope}
{\mathcal{S}}(\Nc) \approx \SlopeOne \exp \left( \sum_{k=1}^{5} B_k
(\ln \Nc)^k \right).
\end{equation}

\noindent here \SlopeOne, $A_k$\ and $B_k$\ are free parameters
obtained by fitting the simulated data, in particular \SlopeOne\
is the slope for $\Nc = 1$.

Since an accuracy of some percent in determining the free
parameters of $\CrFit(\VOT, \Nc)$\ is enough, the fitting
procedure was simplified as follow. For a given compressor, signal
component, swap status, and \Nc\ value $\mathcal{I}$\ and
$\mathcal{S}$\ where determined by a $\chi^2$\ fitting procedure.
The list of $\mathcal{I}$\ and $\mathcal{S}$\ as a function of
\Nc\ have been fitted by using relations
(\ref{eq:cr:fit:intercept}) and (\ref{eq:cr:fit:slope})
respectively. The fitting algorithm tests different degrees of the
polynomial in the aforementioned relations (up to 2 for
$\mathcal{I}(\Nc)$, up to 5 for $\mathcal{S}(\Nc)$)  stopping when
the maximum deviation of the fitted relation respect to the data
is smaller than $0.5\%$\ for $\mathcal{I}$\ or 0.0001 for
$\mathcal{S}$, or when the maximum degree is reached.

Tables \ref{tab:cmpres:30w}, \ref{tab:cmpres:30f},
\ref{tab:cmpres:100w}, \ref{tab:cmpres:100f}\ report the results
of the compression exercise ordered for decreasing \CrOne. The
first column is the name of compression macro (i.e. a given
compression program with a defined selection of modificators and
switches) as listed in tables:
 \ref{tab:press:list:a},
 \ref{tab:press:list:c}.
  The third and fourth column are the fitted
\CrOne and \SlopeOne\ as defined in: (\ref{eq:cr:fit}),
(\ref{eq:cr:fit:intercept}), (\ref{eq:cr:fit:slope}). From the
$5^{\mathrm{th}}$\ to the $7^{\mathrm{th}}$\ columns and from the
$8^{\mathrm{th}}$\ to the $13^{\mathrm{th}}$\ columns the
polynomial degree and the expansion parameters for
(\ref{eq:cr:fit:intercept}) and (\ref{eq:cr:fit:slope}) are
reported.

Many compressors are sensitive to the ordering of the Least and
Most Significant Bytes of a 16 bits word in the computer memory
and files. Two ordering conventions are assumed: {\em UnSwapped}
i.e. Least Significant Byte is stored First or {\em Swapped} i.e.
Most Significant Byte is stored First. As in Digital VAX/VMS
Operating System, Microsoft Windows/NT operating system convention
is Most Significant Byte first. For this reason each test was
repeated twice, one time with the original data stream file with
swapped bytes and the other after unswapping bytes. If the gain in
$\CrOne$\ after unswapping is bigger than some percent, unswapped
compression is reported, otherwise the swapped one is reported.
These two cases are distinct by the second column of
tables \ref{tab:cmpres:30w}, \ref{tab:cmpres:30f},
\ref{tab:cmpres:100w}, \ref{tab:cmpres:100f} which is marked with
a {\em y} if unswapping is applied before compressing. It is
interesting to note that not only 16 bits compressors, such as
{\tt uses}, are sensitive to swapping. Also many 8 bits
compressors are sensitive to it, maybe that this is due to the
fact that if the most probable 8 bits symbol is presented first at
the compressor a slightly better balanced coding table is built.

It should be noted that the coefficients reported here are
obtained compressing one or more full scan circles at a time, so
their use to extrapolate \Cr\ when each scan circle is divided in
small chunks which are separately compressed has to be performed
carefully, especially for $\VOT \approx 0.5$\ V/K\ where some
extrapolated \Cr\ grows instead of to decrease for a decreasing
\Nc\ as in most of the cases. However we did not investigate
further the problem because the time required to perform all the
tests over all the compressors increases decreasing \Nc, and
because up to now a final decision about the packet length has not
been made yet. Moreover, short data chunks introduce other
constrains which are not accounted for by eq. (\ref{eq:cr:theory})
but which are discussed in section \ref{sec:packeting}.

Apart from the choice of the best compressor, Tables
\ref{tab:cmpres:30w}, \ref{tab:cmpres:30f}, \ref{tab:cmpres:100w},
\ref{tab:cmpres:100f}\ allows interesting comparisons.

The performances of the arithmetic compression {\tt
arith} are very sensitive to changes in the coding order $n = 0$,
$\dots$, 7. The computational weight grows with $n$, while \Cr\ is
minimal at $n=0$, maximal for $n=1$\ and decreases increasing $n$
further.

Both non-Adaptive Huffman ({\tt huff-c}) and Adaptive Huffman
({\tt ahuff-c}) are in the list of the worst compressors,
considering both the pure white noise signal and the full signal.

We implemented the space-qualified {\tt uses} compressor with a
wide selection of combinations of its control parameters: the
number of coding bits, the number of samples per block, the
possibility to search for correlations between neighborhood
samples. We report the tests for 16 bits coding only, changing the
other parameters. {\tt Uses} is very sensitive to byte unswapping,
when not performed {\tt uses} does not compress at all. On the
other hand, opposite to {\tt arith}\ the sensitivity of the final
\Cr\ to the various control parameters is small or negligible. In
most cases \CrOne\ differs of less than $0.01$\ for changing the
combination of control parameters, such changes are not displayed
by the two digits approximation in the tables, but they are
accounted for by the sorting procedure which fixes the table
ordering. At 30 GHz most of the tested compressors cluster around $\CrOne =
2.67$\ and at this level {\tt arith-n3}\ is as good as {\tt uses}.
At 100 GHz the best {\tt uses}\ macros clusters around $\CrOne =
2.43$ - $2.44$, equivalent to {\tt arith-n2}\ performances. In our
tests {\tt uses}\ performs worst at 8 samples per block without
correlation search, but apart from it, in our case the correlation
search does not improve significantly the compression
performances. Some commercial programs such as {\tt boa}, {\tt
bzip} compress better than {\tt uses}.

\section{Further Constrains: Packet Independence and Packet
Length}\label{sec:packeting}

As an example of global constrains to the on-board compression we
discuss the problems related to Packets Independence and Packets
Length.

Data from the LFI must be packetized before being sent
to Earth. Packets independence is considered to be a requirement,
then each packet must be self-consistent, its loss or its
erroneous transmission must not interfere with the data retrieval
from subsequent packets. More over each packet must carry in ``clear''
format (i.e. uncompressed) all the information needed to decode
its content. That is: each packet must contain its own decoding table
or decoding information. A typical packet length is about some
hundred of bytes, but smaller length may be planned if required;
at the same time a typical decoding table holds something less
than a hundred bytes leaving limited room for data.

In addition, for a fixed length $\Lu$\ of a random input stream
(expressed in bits) the output $\Lc$\ will not be a constant but
will change in time with respect to the averaged length $\Lu/\Cr$.
Of course, it is not possible to predict in advance what will be
the final length of a given bit stream. So either $\Lu$\ is held
fixed, loosing in compression efficiency, or $\Lu$\ is adapted
with some interactive method, maximizing the compression
efficiency but at the cost of a significant slowing of the
compression process.

In conclusion, the packets independence plus limited packet length
prevents from sending the decoding table, leaving only two
possibilities open: i) send the relevant bytes only
(\cite{Coding:Compression}), ii) to use a predefined coding table
(\cite{Fast:Compression}), both methods are described in the next
section.

\section{Proposed Coding and Compression
Scheme}\label{sec:coding:scheme}

The basic principle of the first method named {\em Least
Significant Bits Packing} (LSBP) is to send only those bits of the
16 bits output from the ADC which are affected by the signal and
the noise. This is effective for the nominal mission since with
the planned quantization step of 0.3 mK/adu, at one sigma the
noise will fill about 21 levels, this will require at least 5 bits
over 16 and it is reasonable to expect a final data flow
equivalent to  $\CrOne < 3$. It is not possible to improve much
the compression rate by compressing the resulting 5 bits data
stream, since its entropy would be $H < 5.4$\ bits and $\Cr \lsim
1.08$.

In order to ensure the compression to be lossless all the samples
exceeding the [$-\sigma$, $+\sigma$]  (5 bits) range have to be
sent separately coding at the same time: their position (address)
in the stream vector and their value. So, for $\Nbits < 16$\ bits
corresponding to a threshold $\xth = 2^{\Nbits}$, each group of
samples stored into a packet is partitioned into two classes
accordingly with their value $x$:

\begin{center}
\begin{tabular}{llll}
 Regular Samples (RS) & $\langle$def$\rangle$ & all those samples for which: & $|x| \leq \xth$, \\
 &&&\\
 Spike Samples (SS)   & $\langle$def$\rangle$ & all those samples for which: & $|x| \geq \xth.$\\
\end{tabular}
\end{center}

The coding process then consists of two main steps: i) to split
the data stream in Regular and Spike Samples preserving the
original ordering in the stream of Regular Samples, ii) to store
(send) the first \Nbits\ bits of the regular samples and, in a
separated area, the 16 bits values and the location in the
original data stream of each Spike Sample, i.e. Spike Samples will
require more space to be stored than regular ones. The decoding
process will be the reverse of this packing process.

In this scheme each packet will be divided into two main areas: the
Regular Samples Area (RSA) which hold the stream of Regular Samples, the Spike
Sample Area (SSA) which hold the stream of Spike Samples, plus a number of fields
which will contain packing parameters such as: the number of samples, the number
of regular samples, the offset, etc.
Since the number of samples in each area will change randomly
it will be not possible to completely fill a packet. The filling process will
leave an empty area in the packet in average smaller than \Nbits.

In \cite{Coding:Compression} a first evaluation for the 30 GHz
channel is given assuming that the signal is composed only of
white noise plus the CMB dipole. As noticed in section
\ref{subsec:results}\ the cosmological dipole affects the compression
efficiency reducing it of a small amount. To deal with it a
possible solution would be to subdivide each data stream in
packets, subtract to each measure of a given packet the
integer average of samples (computed as a 16 bits integer number) and then
compress the residuals. Each integer average will be sent to Earth
together with the related packet where the operation will be
reversed. Since all the numbers are coded as 16 bits integers all
the operations are fully reversible and no round off error occurs.
However it cannot be excluded that the computational cost of such
operation will compensate the gain in \Cr.

Two schemes are proposed to perform the cosmological dipole
self-adaptement. In {\em Scheme A}\ the average of
samples in the packet are subtracted before coding and then sent
separately. In {\em Scheme B}\ \xth\ is varied proportionally to
the dipole contribution. Both of them assumes that the dipole
contribution is about a constant over a packet length. From this
assumption: $L_p \lsim 200$\ samples i.e. $L_p < 512$\ bytes,
since for $L_p > 512$\ bytes the cosmic dipole contribution can
not be considered as a time constant. For larger packets a better
modeling (i.e. more parameters) will be required in order not to
degrade the compression efficiency.

A critical point is to fix the best \xth, i.e. \Nbits, for a given
signal statistics, coding scheme and packet length $L_p$. Even
here \Cr\ grows with the packet length but it does not change
monotously with \xth. An increase in \xth\ (\Nbits) decreases the
number of spike samples, but increases the size of each regular
sample. While the opposite occurs when \xth\ is decreased, and
when $\Nbits<4$\ bits $\Cr < 1$. For both the schemes the
optimality is reached for $\Nbits = 6$\ bits, but {\em Scheme A}\
is better than {\em B}, with: $\Cr(\mbox{{\em Scheme A}}$, $L_p =
512 \mbox{ bytes}) = 2.61$, $\Cr(\mbox{{\em Scheme B}}$, $L_p =
512 \mbox{ bytes}) = 2.29$.

Compared with {\tt arith-n1}, this compression rate is smaller of
about a 14 - 30\%. This is due to two reasons: i) coding by a
threshold cut is less effective than to apply an optimized
compressor; ii) the results reported in tables \ref{tab:cmpres:30w},
\ref{tab:cmpres:30f},
\ref{tab:cmpres:100w},
\ref{tab:cmpres:100f} refer to
the compression of a full circle of data instead
of a small packet, resulting in a higher efficiency. However, the
efficiency of this coding method is similar to the efficiency of
the bulk of the other true loss-less compressors tested up to now,
and when the need to send a decoding table is considered, is even
higher.

The second possible solution to the packeting problem is to use
one or more standardized coding tables for the compression scheme
of choice (\cite{Fast:Compression}). In this case the coding table
would be loaded into the on-board computer before launch or time
by time in flight and the table should be known in advance at
Earth. Major advantages would be: 1. the coding table has not to
be sent to Earth; 2. the compression operator will be reduced to a
mapping operator which may be implement as a tabular search,
driven by the input 8 or 16 bits word to be compressed; 3. any
compression scheme (Huffman, arithmetic, etc.) may be implemented
replacing the coding table without changes to the compression
program; 4. the compression procedure may be easily written in C
or the native assembler language for the on-board computer or,
alternatively, a simple, dedicated hardware may be implemented and
interfaced to the on-board computer. The disadvantages of this
scheme are: 1. each table must reside permanently in the central
computer memory unless a dedicated hardware is interfaced to
it; 2. it is difficult to use adaptive schemes in order to tune
the compressor to the input signal, as a consequence the \Cr\ may
be somewhat smaller than in the case of a true self-adapting
compressor code.

The first problem may be circumvented limiting the length of the
words to be compressed. In our case the data streams may be
divided in chunks of 8 bits and the typical table size would be
$\lsim 1$\ Kbyte. Precomputed coding tables may be accurately
optimized by Monte-Carlo simulations on ground or using signals
from ground tests of true hardware.

The second problem may be overcome by using a preconditioning
stage, reducing the statistics of the input signal to the
statistics for which the pre-calculated table is optimized. In
addition more tables may reside in the computer memory and
selected looking to the signal statistics. With a simple
reversible statistical preconditioner, about ten tables
per frequency channel would be stored in the computer memory, so
that the total memory occupation would be less than about 40
Kbytes. It cannot be excluded that the two methods just outlined
cannot be merged.

\section{Estimation of the Overall Compression
Rate}\label{sec:overall:cr}

The overall compression rate (efficiency) is the average of \Cr\
(\etac) over the full set of detectors. Appendix \ref{appendix:a}\
illustrates the mathematical aspects of such average. From
(\ref{eq:overall:etac}):

\begin{equation}
 \overline{\Cr}(\Nc) =
  \left[\sum_\nu \frac{f_\nu}{\Cr_{,\nu}(\Nc)}\right]^{-1}.
\end{equation}

\noindent We will limit ourselves to the most probable case $\Nc =
1$\ and to the most effective compressor {\tt arith-n1}. The
compression parameters \CrOne\ and \SlopeOne\ at 30 GHz and
100 GHz are derived from our simulations, while \CrOne\ and
\SlopeOne\ at 44 GHz and 70 GHz are obtained by linear
interpolation of the simulated values as a function of $\ln
\sigma_\nu$. After that we obtain:

\begin{equation}
 \overline{\Cr} \approx \frac{2.66}{1+0.271 \times \ln \VOT}.
\end{equation}

\noindent
 As expected the overall compression rate is dominated by the
100 GHz channel. Taking in account the conservative \VOT\
distribution considered in equation
(\ref{eq:average:etac:approx}) the overall compression rate
becomes: $\overline{\Cr} \approx 2.63$\ which represents a
$\approx 2 \%$\ correction only. It is likely that this correction
will be even smaller, since the amplifiers gain will be adjusted
in order to cover a smaller \VOT\ interval. So this $2\%$\
correction represents our greatest uncertainty in our estimation
of the expected compression rate, and we may conservatively
conclude that:

\begin{equation}
 \overline{\Cr}_{,\mathrm{arith-n1}} \approx 2.65 \pm 0.02
\end{equation}

\noindent Recently a new evaluation of the expected instrumental
sensitivity leads to some change in the expected white noise
r.m.s.. These changes affect in particular the 30 GHz channel, but
does not change significantly the 100 GHz channel so that the
overall compression rate will be practically unaffected.

\section{Conclusions}\label{sec:conclusions}

The expected data rate from the Planck Low Frequency Instrument is $\approx
260$\ kbits/sec. The bandwidth for the scientific data download
currently allocated is just $\approx 60$\ kbit/sec. Assuming an
equal subdivision of the bandwidth between the two instruments
on-board Planck, an overall compression rate of a factor 8.7 is
required to download all the data.

In this work we perform a full analysis on realistically simulated
data streams for the 30 GHz and 100 GHz channels in order to fix
the maximum compression rate achievable by loss-less compression
methods, without considering explicitly other constrains such as:
the power of the on-board Data Processing Unit, or the
requirements about packet length limits and independence, but
taking in account all the instrumental features relevant to data
acquisition, i.e.: the quantization process, the temperature /
voltage conversion, number of quantization bits and signal
composition.

As a complement to the experimental analysis we perform in
parallel a theoretical analysis of the maximum compression rate.
Such analysis is based on the statistical properties of the
simulated signal and is able to explain quantitatively most of the
experimental results.

Our conclusions about the statistical analysis of the quantized
signal are:
 I) the nominally quantized signal has an entropy $h \approx 5.5$\
bits at 30GHz and $h \approx 5.9$\ bits at 100GHz, which allows a
theoretical upper limit for the compression rate $\approx 2.9$\ at 30 Ghz
and $\approx 2.7$\ at 100 GHz.
 II) Quantization may introduce some distortion in the signal statistics
but the subject requires a deepest analysis.

Our conclusions about the compression rate are summarized as
follows:
 I) the compression rate \Cr\ is affected by the quantization step, since greater is the
quantization step higher is \Cr\ (but worse is the measure
accuracy).
 II) \Cr\ is affected also by the stream length $L_u$, i.e. more circles
are compressed better then few circles.
 III) the dependencies on the quantization step and $L_u$\ for each compressor may be
 summarized by the empirical formula (\ref{eq:cr:fit}). A reduced
 compression rate $\CrOne$\ is correspondingly defined.
 IV) the \Cr\ is affected by the signal composition, in particular,
by the white noise r.m.s. and by the dipole contribution, the
former being the dominant parameter and the latter influencing
\Cr\ for less than $\approx 6\%$.  The inclusion of the dipole
contribution reduces the overall compression rate. The other
components (1/f noise, CMB fluctuations, the galaxy, extragalactic
sources) have little or no effect on \Cr. In conclusion, for the
sake of compression rate estimation, the signal may be safely
represented by a sinusoidal signal plus white noise.
 V) since the noise r.m.s. increases with the frequency, the
compression rate \Cr\ decreases with the frequency, for the LFI
$\Delta\Cr/\Cr \lsim 10\%$.
 VI) the expected random r.m.s. in the overall compression rate is less than
$1\%$.
 VII) we tested a large number of off-the-shelf compressors, with
many combinations of control parameters so to cover every
conceivable compression method.
The best performing compressor is the arithmetic
compression scheme of order 1: {\tt arith-n1}, the final \CrOne\
being 2.83 at 30 GHz and 2.61 at 100 GHz. This is significantly
less than the bare theoretical compression rate
(\ref{eq:cr:theory}) but when the quantization process is taken
properly into account in the theoretical analysis, this
discrepancy is largely reduced.
 VIII) taking into account the data flow distribution
among different compressors the overall compression rate for {\tt
arith-n1} is:

$$
 \overline{\Cr}_{,\mathrm{arith-n1}} \approx 2.65 \pm 0.02
$$

\noindent This result is due to the nature of the signal which is
noise dominated and clearly excludes the possibility to reach the
required data flow reduction through loss-less compression only.

Possible solutions deal with the application of lossy compression
methods such as: on-board averaging, data rebinning, or averaging
of signals from duplicated detectors, in order to reach an overall
lossy compression of about a factor 3.4, which coupled with the
overall loss-less compression rate of about 2.65 should allow to
reach the required final compression rate $\approx 8.7$. However
each of these solutions will introduce heavy constraints and
important reduction of performances in the final mission design,
so that careful and deep studies will be required in order to
choose the best one.

Another solution to the bandwidth problem would be to apply a
coarser quantization step. This has however the drawback of
reducing the signal resolution in terms of $\Delta T/T$.

Lastly the choice of a given compressor cannot be based only on
its efficiency obtained from simulated data, but also on the
on-board available CPU and on the official ESA space
qualification: tests with this hardware platform and other
compressors will be made during the project development. Moreover,
in the near future long duration flight balloon experiments and
ground experiments (see \cite{Lasenby:etal:1998,Debe:masi:98})
will provide a solid base to test and improve compression
algorithms. In addition the final compression scheme will have to
cope with requirements about packet length and packet
independence. We discuss briefly this problems recalling two
proposals (\cite{Fast:Compression}, \cite{Coding:Compression}) which
suggest solutions to cope with these constrains.

\appendix
\section{Appendix: Formulation of the Final Data
Flow}\label{appendix:a}

In this appendix we will discuss how to account for the
distribution of the acquisition parameters between the different
detectors in the computation of the overall compression rate.
Since the formalism is simpler we will develop expressions for
$\etac = 1/\Cr$\ instead of \Cr.

We have pointed out in \ref{sec:Entropy}\ that the compression
efficiency is a random variable, whose distribution is a function
of all those parameters which are relevant to fix the statistical
distribution of the input signal. In our case: $\nu$, \VOT, \AFO,
\NCirc\ are the relevant parameters, so that the conditioned
probability to have a compression efficiency in the range $\etac$,
$\etac + d\etac$\ is:

\begin{equation}
  {\mathcal P}_{\nu,N_{\mathrm{Circ}}}\left(\etac | \AFO, \VOT \right) d
 \etac.
\end{equation}

\noindent This probability may be obtained by our MonteCarlo
simulations for different combinations of \AFO, \VOT, \NCirc\ and
$\nu$. Then the averaged compression efficiency is:

\begin{equation}
 \overline{\eta}_{\mathrm{{c}}\nu,N_{\mathrm{Circ}}}(\AFO, \VOT) = \int_0^{+\infty} d\etac\,
 \etac\,
 {\mathcal P}_{\nu,N_{\mathrm{Circ}}}
 \left(\etac | \AFO, \VOT\right).
\end{equation}

\noindent Of course we assumed that for any $\nu$, \VOT, \AFO,
\NCirc\ the probability distribution is integrable and normalized
to 1, while the integration limits $0$, $+\infty$\ are to be
intended as formal. There are several detectors for any
frequency channel, each one having its own \AFO\ and \VOT, so
distributions of \AFO\ and \VOT\ values may be guessed among the
different detectors. Assuming they are integrable and normalized
to 1 as well it is possible to compute the most probable
$\aetac_{\nu,N_{\mathrm{Circ}}}$\ as \footnote{Here
 $$\int_{\mathrm{AFO}_{\mathrm{min}}}^{\mathrm{AFO}_{\mathrm{max}}} d\mathrm{AFO} \,
{\mathcal P}_\nu(\AFO) = 1, \; \; \;
\int_{\mathrm{VOT}_{\mathrm{min}}}^{\mathrm{VOT}_{\mathrm{max}}}
d\mathrm{VOT} \, {\mathcal P}_\nu(\mathrm{VOT}) = 1$$. };

\begin{equation}\label{eq:average:etac}
 \aetac_{\nu,N_{\mathrm{Circ}}} =
  \int_{\mathrm{AFO}_{\mathrm{min}}}^{\mathrm{AFO}_{\mathrm{max}}} d\AFO \, {\mathcal P}_\nu(\AFO) \;
   \int_{\mathrm{VOT}_{\mathrm{min}}}^{\mathrm{VOT}_{\mathrm{max}}} d\VOT \, {\mathcal P}_\nu(\VOT) \;
  \aetac_{\nu,N_{\mathrm{Circ}}}(\AFO, \VOT).
\end{equation}

\noindent With this definition the final overall compression
efficiency is:

\begin{equation}\label{eq:overall:etac}
  \aetac_{N_{\mathrm{Circ}}} = \sum_{\nu = 30, 44, 70, 100 \mathrm{GHz}}
                    f_\nu
                    \aetac_{\nu,N_{\mathrm{Circ}}}
\end{equation}

\noindent where $f_\nu$\ is the partition function for the data
flow through the different detectors, if $n_{\mathrm{dtc},\nu}$\
is the number of detectors for the frequency channel $\nu$\ (see
Tab. I), $n_{\mathrm{dtc}} =   \sum_{\nu = 30, 44, 70, 100
\mathrm{GHz}} n_{\mathrm{dtc},\nu} = 112$, is the total number of
detectors and if the number of samples for frequency is a
constant, then:

\begin{equation}
  f_\nu = \frac{n_{\mathrm{dtc},\nu}}{112},
\end{equation}

\noindent so that for $\nu = 30$, $44$, $70$\ and 100 GHz\
respectively: $f_\nu = 0.0714$, $0.1071$, $0.2143$\ and $0.6071$,
finally the expect data rate for each set of 60 circles is:

\begin{equation}
  \overline{R}_{N_{\mathrm{Circ}}} =
  16 \, \mathrm{bits} \; \times \;
  60 \, \mathrm{circles} \; \times \;
  8640 \, \mathrm{samples} \; \times \;
  112 \, \mathrm{detectors} \; \times \;
  \aetac^{N_{\mathrm{Circ}}}.
\end{equation}

\noindent Presently there are no data to know in advance the
distribution of \VOT\ and \AFO\ values between the different
detectors. For this reason in this work we assumed simply flat
distributions, identical for each frequency for such parameters.
More over, the \AFO\ contribution is negligible, so that the
variance introduced by this parameter is neglected. From
(\ref{eq:cr:theory}) we assumed that the compression efficiency is
approximately a linear function of $\ln\VOT$\ or:

\begin{equation}
 \aetac_{\nu,N_{\mathrm{Circ}}}(\VOT) \approx
 \aetac_{\nu,N_{\mathrm{Circ}},1} +
 \daetac_{\nu,N_{\mathrm{Circ}}} \ln \VOT
\end{equation}

\noindent where $\daetac_{\nu,N_{\mathrm{Circ}}}$\ is the first
derivative of $\aetac_{\nu,N_{\mathrm{Circ}}}(\VOT)$\ with respect
to $\ln\VOT$\ computed for $\VOT=1$\ V/K,
$\aetac_{\nu,N_{\mathrm{Circ}},1} \equiv
\aetac_{\nu,N_{\mathrm{Circ}}}(\VOT = 1\;\mathrm{V/K})$. As an
example, at 30 GHz for {\tt arith-n1}\ the
full signal compression rate is
$\aetac_{\nu,N_{\mathrm{Circ}}}(\VOT) \approx 0.3534  + 0.287
\times \ln \VOT(\mathrm{K/V})$\ with one interpolation error less
than $0.2\%$. With these approximations eq.
(\ref{eq:average:etac}) becomes

\begin{equation}\label{eq:average:etac:approx}
 \aetac_{\nu,N_{\mathrm{Circ}}} \approx
 \aetac_{\nu,N_{\mathrm{Circ}},1} +
  \daetac_{\nu,N_{\mathrm{Circ}}}
  \int_{0.5\;\mathrm{V/K}}^{1.5\;\mathrm{V/K}} d\VOT \,
  \frac{ \ln \VOT}{1.0 \;\mathrm{V/K}}
\end{equation}

\noindent and after integration we obtain the final formula

\begin{equation}\label{eq:average:etac:final}
 \aetac_{\nu,N_{\mathrm{Circ}}} \approx
 \aetac_{\nu,N_{\mathrm{Circ}},1} - 0.045229 \cdot
\daetac_{\nu,N_{\mathrm{Circ}}}
\end{equation}

\noindent for the case in the previous example:
$\aetac_{\nu,N_{\mathrm{Circ}}=2} \approx 0.3404$\ which is
equivalent to a compression efficiency $\approx 2.94$.

To understand the influence of the error in the \VOT\
determination over the distribution on the final predictions the
computation is made for a truncated (i.e. zero outside the \VOT\
range of interest) normal distribution of \VOT. The r.m.s. for the
\VOT\ distribution is chosen in the \VOT\ range [0.5, 1.5]\ V/K we
obtain respectively $\aetac_{\nu,N_{\mathrm{Circ}}} \approx
0.3494$, $0.3439$, $0.3420$; which corresponds to compression
efficiencies: 2.86, 2.91, 2.92 respectively. Similar results are
obtained with a quadratic \VOT\ distribution. In conclusion these
predictions are robust against the shape of the \VOT\
distribution, at least for distributions which are symmetric
around the nominal $\VOT = 1$\ V/K value.

\acknowledgments{
We warmly acknowledge a number of people which actively support
this work with fruitful discussions, in particular
 F. Arg\"ueso,
 M. Bersanelli,
 L. Danese,
 G. De~Zotti,
 E. Gazt\~naga,
 J. Herrera,
 N. Mandolesi,
 P. Platania,
 A. Romeo,
 M. Seiffert and
 L. Toffolatti
and
 K. Gorski and all people involved in the construction
of the Healpix pixelisation tools, largely employed in this work,
and Dr. G. Lombardi from Siemens Bochun -
Germany and Dr. G. Maris from {\em ETNOTEAM - Milano} for fruitful
discussions about compression principles and their practical
application.
}

%

\newpage

\begin{table}[t]
\caption{Summary of LFI characteristics as recently revised by the
LFI Consortium (\cite{Mandolesi:etal:1999}). Data rates are
tabulated for the case of a sampling rate equal to 8640 samples
per circle and constant time and frequency.}\label{tab:PlanckTable}
\begin{center}
\renewcommand{\arraystretch}{0.8}
\begin{tabular}{lrrrr}
\hline\hline
 Center frequency $\nu$ [GHz]          & 30 & 44 & 70 & 100 \\
 Number of detectors $n_{\mathrm{dtc},\nu}$ & 8  & 12  & 24 & 68  \\
 Angular resolutions, FWHM [$'$] & 33.6 & 22.9 & 14.4 & 10.0  \\
 Bandwidth [$\Delta \nu / \nu$]  & 0.2 & 0.2 & 0.2 & 0.2 \\
 $10^6 \Delta T / T$             & 1.6 & 2.4 & 3.6 & 4.3 \\
 $\Delta T_{ant}$ [$\mu$K]       & 5.1   & 7.8   & 10.6  & 12.4  \\
 $\Delta T_{ant}$ [mK] per sampling and receiver
            & 2.06  & 2.61 & 3.16 & 4.36 \\
 Number of samples for beam & 13.4  & 9.2 & 5.8 & 4.0 \\
 Data rate for detector [Kb/sec]  &  2.3 & 2.3  & 2.3 & 2.3 \\
 Data rate for frequency [Kb/sec] & 18.4 & 27.6 & 55.3 & 156.7 \\
 Uncompressed data rate partition function $f_\nu$\ [$\%$] & 7.14 & 10.71 & 21.43 & 60.71 \\
\hline
\end{tabular}
\renewcommand{\arraystretch}{1}
\end{center}
\end{table}

 \begin{table}
 \caption{Entropy for 16 bits samples at 30 and 100 GHz, for
only White Noise and Full Signal as a function of \Lchunck. Total
Entropy refers to the entropy computed over the full set of
samples ($8640 \times 60$), Mean and RMS Entropy are the mean and
RMS of different realizations of chunks of samples of length
\Lchunck. The same for \Cr\ columns. Here \Cr\ are derived from
the corresponding values of the entropy. The quantization step is
$\Delta = 0.305$\ mK/adu.
 }\label{tab:entropy}
 \begin{center}
 \renewcommand{\arraystretch}{0.6}
 \begin{tabular}{rllllll}
 \multicolumn{7}{c}{30 GHz, White Noise}\\
 \hline
 &
 \multicolumn{3}{c}{Entropy (bits)}&
 \multicolumn{3}{c}{\Cr}\\
 \multicolumn{1}{c}{\Lchunck}  &
 \multicolumn{1}{c}{Total}  &
 \multicolumn{1}{c}{Mean}  &
 \multicolumn{1}{c}{RMS}  &
 \multicolumn{1}{c}{Total}  &
 \multicolumn{1}{c}{Mean}  &
 \multicolumn{1}{c}{RMS}  \\
 \hline\hline
    16 & 5.1618 & 3.5596 & 0.1989 & 3.10    & 4.49 & 0.251 \\
    32 & 5.1618 & 4.1815 & 0.1658 & 3.10    & 3.83 & 0.152 \\
    64 & 5.1618 & 4.6108 & 0.1262 & 3.10    & 3.47 & 0.095 \\
   135 & 5.1618 & 4.8791 & 0.0890 & 3.10    & 3.28 & 0.060 \\
  8640 & 5.1618 & 5.1561 & 0.0114 & 3.10    & 3.10 & 0.007 \\
 17280 & 5.1618 & 5.1589 & 0.0061 & 3.10    & 3.10 & 0.004 \\
 \hline
 %
 %
 &&&&&&\\
 \multicolumn{7}{c}{30 GHz, Full Signal}\\
 \hline
 &
 \multicolumn{3}{c}{Entropy (bits)}&
 \multicolumn{3}{c}{\Cr}\\
 \multicolumn{1}{c}{\Lchunck}  &
 \multicolumn{1}{c}{Total}  &
 \multicolumn{1}{c}{Mean}  &
 \multicolumn{1}{c}{RMS}  &
 \multicolumn{1}{c}{Total}  &
 \multicolumn{1}{c}{Mean}  &
 \multicolumn{1}{c}{RMS}  \\
 \hline\hline
    16 & 5.5213 & 3.5602 & 0.1982 & 2.90    & 4.49 & 0.250 \\
    32 & 5.5213 & 4.1849 & 0.1664 & 2.90    & 3.82 & 0.152 \\
    64 & 5.5213 & 4.6162 & 0.1278 & 2.90    & 3.47 & 0.096 \\
   135 & 5.5213 & 4.8885 & 0.0893 & 2.90    & 3.27 & 0.060 \\
  8640 & 5.5213 & 5.5119 & 0.0176 & 2.90    & 2.90 & 0.009 \\
 17280 & 5.5213 & 5.5157 & 0.0118 & 2.90    & 2.90 & 0.006 \\
 \hline
 %
 %
 &&&&&&\\
 \multicolumn{7}{c}{100 GHz, White Noise }\\
 \hline
 &
 \multicolumn{3}{c}{Entropy (bits)}&
 \multicolumn{3}{c}{\Cr}\\
 \multicolumn{1}{c}{\Lchunck}  &
 \multicolumn{1}{c}{Total}  &
 \multicolumn{1}{c}{Mean}  &
 \multicolumn{1}{c}{RMS}  &
 \multicolumn{1}{c}{Total}  &
 \multicolumn{1}{c}{Mean}  &
 \multicolumn{1}{c}{RMS}  \\
 \hline\hline
    16 & 5.7436 & 3.6962 & 0.1740 & 2.79    & 4.33 & 0.204 \\
    32 & 5.7436 & 4.4174 & 0.1521 & 2.79    & 3.62 & 0.125 \\
    64 & 5.7436 & 4.9627 & 0.1230 & 2.79    & 3.22 & 0.080 \\
   135 & 5.7436 & 5.3354 & 0.0875 & 2.79    & 3.00 & 0.049 \\
  8640 & 5.7436 & 5.7352 & 0.0115 & 2.79    & 2.79 & 0.006 \\
 17280 & 5.7436 & 5.7394 & 0.0063 & 2.79    & 2.79 & 0.003 \\
 \hline
 %
 %
 &&&&&&\\
 \multicolumn{7}{c}{100 GHz, Full Signal }\\
 \hline
 &
 \multicolumn{3}{c}{Entropy (bits)}&
 \multicolumn{3}{c}{\Cr}\\
 \multicolumn{1}{c}{\Lchunck}  &
 \multicolumn{1}{c}{Total}  &
 \multicolumn{1}{c}{Mean}  &
 \multicolumn{1}{c}{RMS}  &
 \multicolumn{1}{c}{Total}  &
 \multicolumn{1}{c}{Mean}  &
 \multicolumn{1}{c}{RMS}  \\
 \hline\hline
    16 & 5.8737 & 3.6970 & 0.1734 & 2.72    & 4.33 & 0.203 \\
    32 & 5.8737 & 4.4186 & 0.1526 & 2.72    & 3.62 & 0.125 \\
    64 & 5.8737 & 4.9655 & 0.1224 & 2.72    & 3.22 & 0.079 \\
   135 & 5.8737 & 5.3419 & 0.0887 & 2.72    & 3.00 & 0.050 \\
  8640 & 5.8737 & 5.8604 & 0.0180 & 2.72    & 2.73 & 0.008 \\
 17280 & 5.8737 & 5.8655 & 0.0127 & 2.72    & 2.73 & 0.006 \\
 \hline
 \end{tabular}
 \renewcommand{\arraystretch}{1}
 \end{center}
 \vspace{0.3cm}
 \end{table}

 \begin{table}
 \caption{Quantization Effect on the Kolmogorow - Smirnov D test applied
to simulated data, $\Delta$\ is the quantization step.
}\label{tab:dtest}
 \begin{center}
 \begin{tabular}{cccc}
    & \multicolumn{3}{c}{$\Delta$\ (mK/adu)} \\
        & 1.220     & 0.610     & 0.406 \\
 \hline
  ${\mathcal{F}}(D < 0.1475, \; \mathrm{White\;Noise})$ & 0.28  & 0.70  & 0.84  \\
  ${\mathcal{F}}(D < 0.1475, \; \mathrm{Signal})$ & 0.27  & 0.71  & 0.86  \\
  $D_{\mathrm{95}}^{\mathrm{Q}} $ & 0.2449  & 0.1851 & 0.1678 \\
  ${\mathcal{F}}(D < D_{\mathrm{95}}^{\mathrm{Q}}, \; \mathrm{Signal})$ & 0.95    & 0.95  & 0.95    \\
 \hline
 \end{tabular}
 \end{center}
 \vspace{0.3cm}
 \end{table}

\begin{center}
\def\IDEM{\multicolumn{1}{c}{"}}
\begin{table}[t]
\caption{Tested compressors and related parameters. The {\em
Macro} column contains the names of the macros running a given
compression {\em Code} with a given combination of {\em
Parameters}}\label{tab:press:list:a}
 \renewcommand{\arraystretch}{0.7}
\begin{tabular}{llll} &&&\\
 \hline
 \hline
 Macro       & Code      & Parameters & Note\\
 \hline
 ahuff-c    & ahuff-c   &             & Adaptive Huffman \cite{Compression:Book}\\
 \hline
 AR         & ar        &             & \\
 \hline
 arc        & arc       &             & \\
 \hline
 arha       & arhangel  &             & http://www.geocities.com/SiliconValley/Lab/6606\\
 arhaASC    & \IDEM     & -1          & ASC method\\
 arhaHSC    & \IDEM     & -2          & HSC method\\
 \hline
 arith-c    & arith-c   &             & Arithmetic coding \cite{Compression:Book}\\
 \hline
 arith-n    & arith-n   &             & Adaptive Arithmetic Coding (AC) \cite{Compression:Book}\\
 arith-n0   & \IDEM     & -o 0        & Zeroth order Arithmetic coding \\
 arith-n1   & \IDEM     & -o 1        & First order AC\\
 arith-n2   & \IDEM     & -o 2        & Second order AC\\
 arith-n3   & \IDEM     & -o 3        & Third order AC\\
 arith-n4   & \IDEM     & -o 4        & Fourth order AC\\
 arith-n5   & \IDEM     & -o 5        & Fifth order AC\\
 arith-n6   & \IDEM     & -o 6        & Sixth order AC\\
 arith-n7   & \IDEM     & -o 7        & Seventh order AC\\
 \hline
 arj        & arj       &             & \\
 arj0       & \IDEM     & -m 0        & method 0 (no compression)\\
 arj1       & \IDEM     & -m 1        & method 1 \\
 arj2       & \IDEM     & -m 2        & method 2 \\
 arj3       & \IDEM     & -m 3        & method 3 \\
 arj4       & \IDEM     & -m 4        & method 4 \\
%
 \hline
 boa        & boa       &             & \\
 \hline
 bzip       & bzip2090  &             & \\
 bziprb     & \IDEM     & --repetitive-best & best compression of repetitive blocks \\
 bziprf     & \IDEM     & --repetitive-fast & fast compression of repetitive blocks \\
 \hline
 gzip1      & gzip      & -1          & fast compression \\
 gzip9      & \IDEM     & -9          & best compression \\
 \hline
 huff-c     & huff-c    &             & Hauffman \cite{Compression:Book}\\
 \hline
 jar        & jar32     &             & \\
 jar1       & \IDEM     & -m1         & method 1\\
 jar2       & \IDEM     & -m2         & method 2\\
 jar3       & \IDEM     & -m3         & method 3\\
 jar4       & \IDEM     & -m4         & method 4\\
 \hline
 lha        & lha       &             & \\
 \hline
 lzss       & lzss      &             & \\
 \hline
 lzw12      & lzw12     &             & \\
 \hline
 lzw15v     & lzw15v    &             & \\
 \hline
\end{tabular}
 \renewcommand{\arraystretch}{1}
\end{table}

\begin{table}[t]
\caption{See table
\ref{tab:press:list:a}.}\label{tab:press:list:c}
 \renewcommand{\arraystretch}{0.8}
\begin{tabular}{llll} &&&\\
 \hline
 \hline
 Macro       & Code      & Parameters & Note\\
 \hline
 pkzip      & pkzip     &             & from PKWARE \\
 pkzip-ef   & \IDEM     & -ef         & fast compression \\
 pkzip-en   & \IDEM     & -en         & normal compression \\
 pkzip-es   & \IDEM     & -es         & super fast compression \\
 pkzip-ex   & \IDEM     & -ex         & extra compression \\
 \hline
 rar-m0     & rar       & -m0         & level 0 compression \\
 rar-m1     & \IDEM     & -m1         & level 1 compression \\
 rar-m2     & \IDEM     & -m2         & level 2 compression \\
 rar-m3     & \IDEM     & -m3         & level 3 compression \\
 rar-m4     & \IDEM     & -m4         & level 4 compression \\
 rar-m5     & \IDEM     & -m5         & level 5 compression \\
 \hline
 splint     & splint    &             & \\
 \hline
 SZIP00     & szip      &             & Rice Algorithm and Rice compression chip simulator\\
 szip0ec    & \IDEM     & -ec         & entropy coding compression mode\\
 szip0nu    & \IDEM     & -nn         & nearest neighbor compression mode\\
 szipc0     & \IDEM     & -chip       & compress exactly as chip\\
 SZIPCEC    & \IDEM     & -chip -ec   & as szip0ec + chip compression \\
 SZIPCNU    & \IDEM     & -chip -nn   & as szip0nu + chip compression \\
 \hline
 uses       & uses      & -n 16 -s 64 -rr & Universal Source Encoder for Space\\
            &           &                 & 16 bits per sample,\\
            &           &                 & 64 samples for scanline,\\
            &           &                 & correlates near samples (CNS) \\
 uses008    & \IDEM     & -n 16 -s 8 -j 8 & 8 samples, 8 samples per block\\
 uses008rr  & \IDEM     & -n 16 -s 8 -rr -j 8 & as uses008 + CNS\\
 uses016    & \IDEM     & -n 16 -s 16     & 16 samples per block\\
 uses016rr  & \IDEM     & -n 16 -s 16 -rr & 16 samples per block + CNS\\
 uses032    & \IDEM     & -n 16 -s 32     & 32 samples per block\\
 uses032rr  & \IDEM     & -n 16 -s 32 -rr & 32 samples per block + CNS\\
 uses064    & \IDEM     & -n 16 -s 64     & 64 samples per block\\
 uses064rr  & \IDEM     & -n 16 -s 64 -rr & 64 samples per block + CNS\\
 uses320    & \IDEM     & -n 16 -s 320    & 320 samples per block\\
 uses320rr  & \IDEM     & -n 16 -s 320 -rr & 320 samples per block + CNS\\
 uses960    & \IDEM     & -n 16 -s 960    & 960 samples per block\\
 uses960rr  & \IDEM     & -n 16 -s 960 -rr & 960 samples per block + CNS\\
 \hline
 zoo        & zoo       &                 & \\
 \hline
\end{tabular}
 \renewcommand{\arraystretch}{1}
\end{table}
\end{center}

 \renewcommand{\arraystretch}{0.5}
\begin{table}[t]
 \caption{Compression Rates at 30 GHz, white noise only}\label{tab:cmpres:30w}
    \vspace{1cm}\mbox{\hspace{-1.3cm}\epsfbox{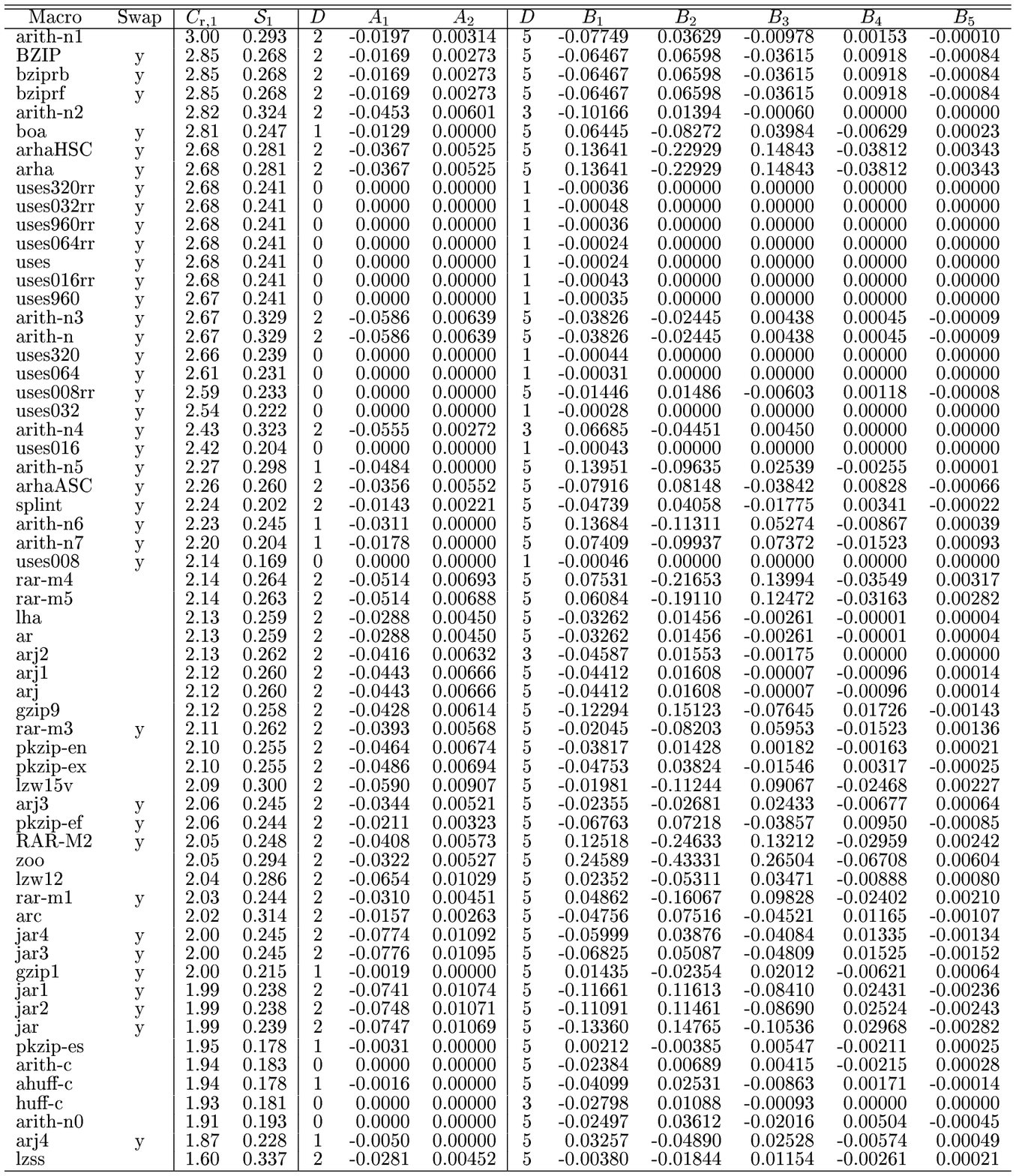}}
\end{table}

\begin{table}[t]
 \caption{Compression Rates at 30 GHz, full signal}\label{tab:cmpres:30f}
    \vspace{1cm}\mbox{\hspace{-1.3cm}\epsfbox{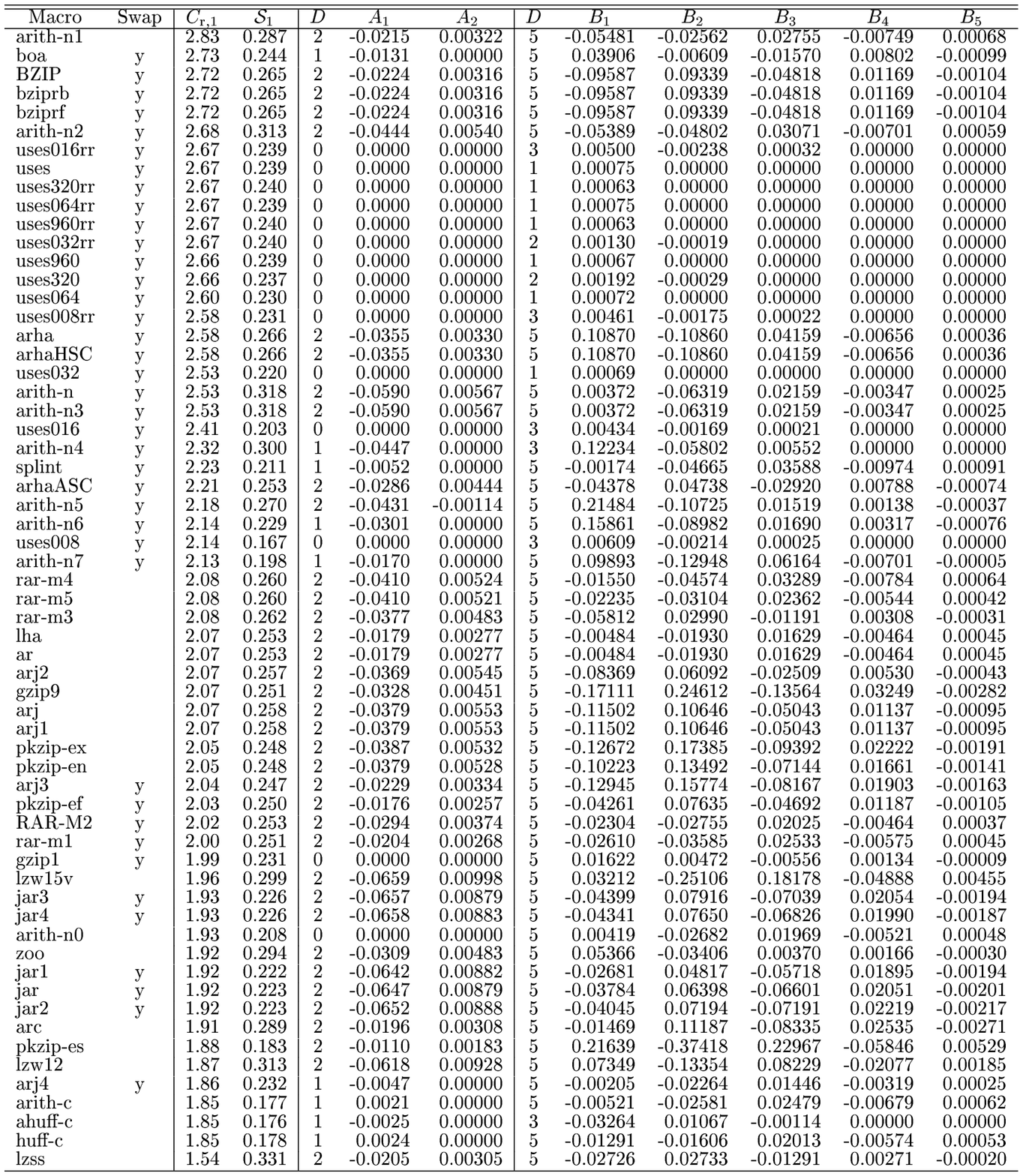}}
\end{table}

\begin{table}[t]
 \caption{Compression Rates at 100 GHz, white noise only}\label{tab:cmpres:100w}
    \vspace{1cm}\mbox{\hspace{-1.3cm}\epsfbox{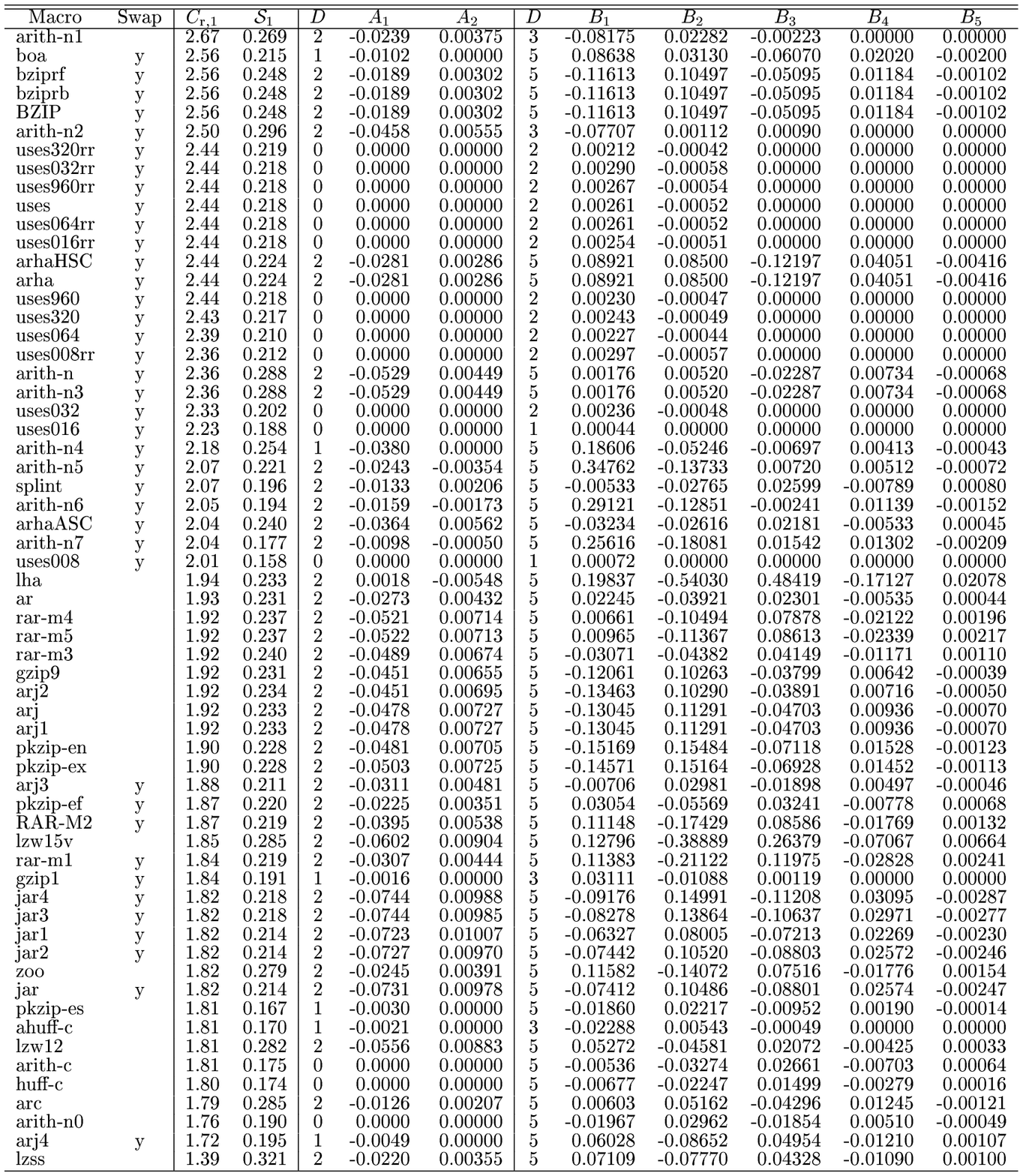}}
\end{table}

\begin{table}[t]
  \caption{Compression Rates at 100 GHz, full signal}\label{tab:cmpres:100f}
    \vspace{1cm}\mbox{\hspace{-1.3cm}\epsfbox{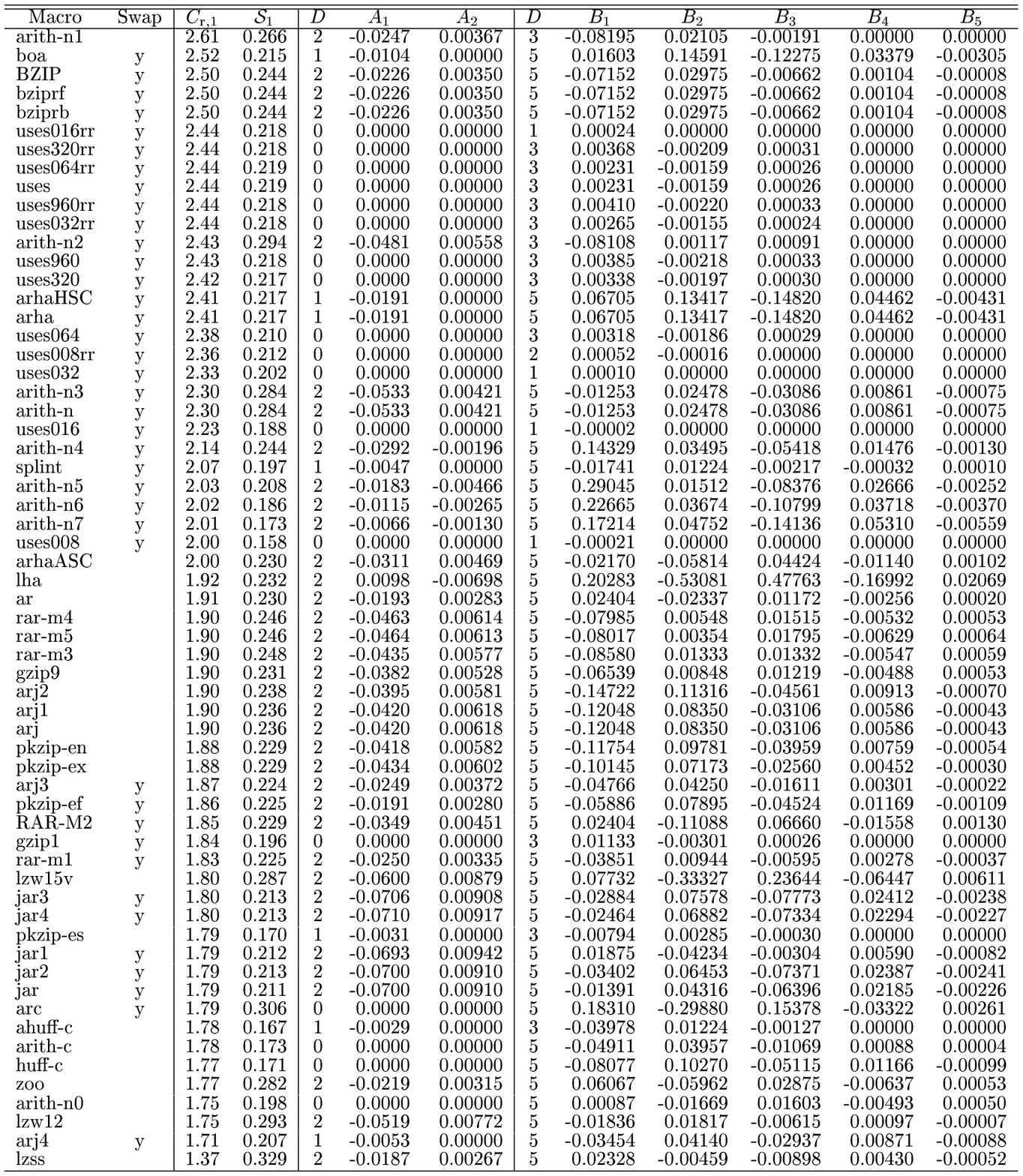}}
\end{table}
 \renewcommand{\arraystretch}{1}


\newpage

\begin{figure}[t]\begin{center}
 \epsfysize=15cm\epsfbox{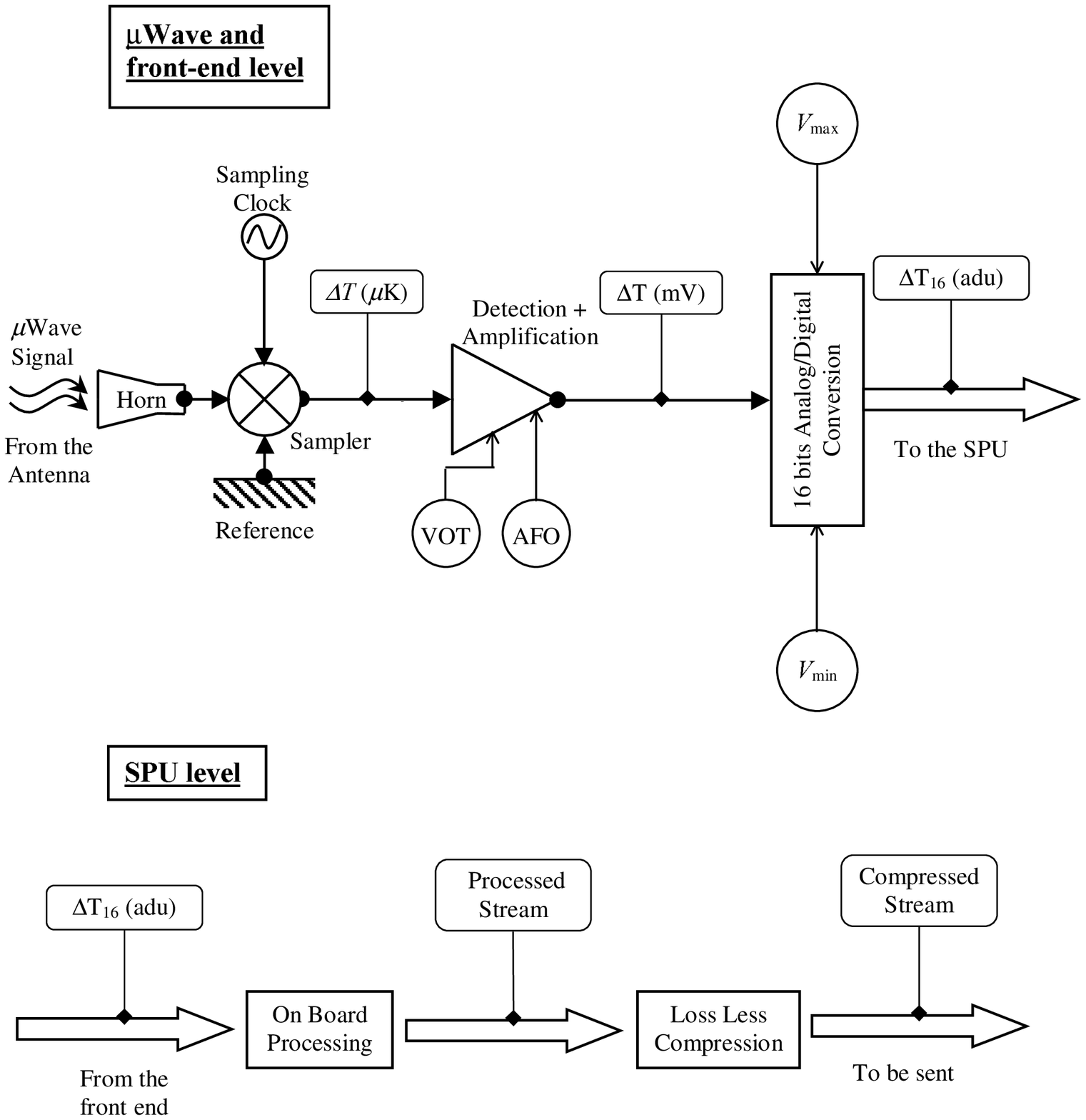}
 \caption[h]{
 Scheme for the functional model of the acquisition pipeline.
 }\label{fig:acquisition:pipe}
\end{center} \end{figure}

\begin{figure}[t]\begin{center}
 \epsfysize=7cm\epsfbox{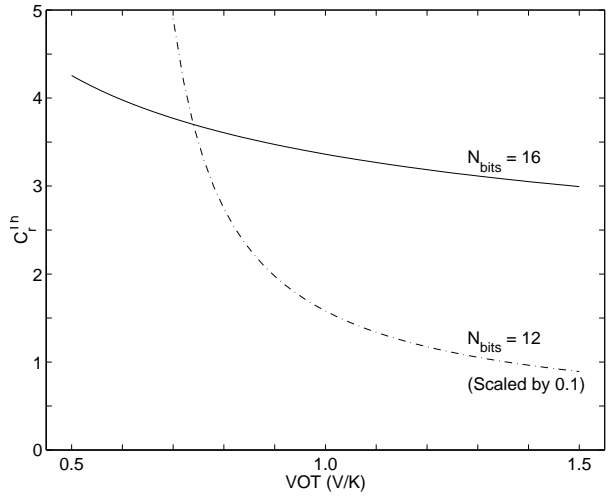}
 \caption[h]{
 \CrTh\ as a function of \VOT\ and \Nbits. It is
 assumed $\Vmin = -10$\ V, $\Vmax = +10$\ V and $\sigma_l = 2\times
 10^{-3}$\ K. The curve for 12 bits is scaled by a factor 0.1 to
 allow a better comparison with the 16 bits curve.
 }\label{fig:CrTh}
\end{center} \end{figure}

\begin{figure}[t]\begin{center}
 \plotone{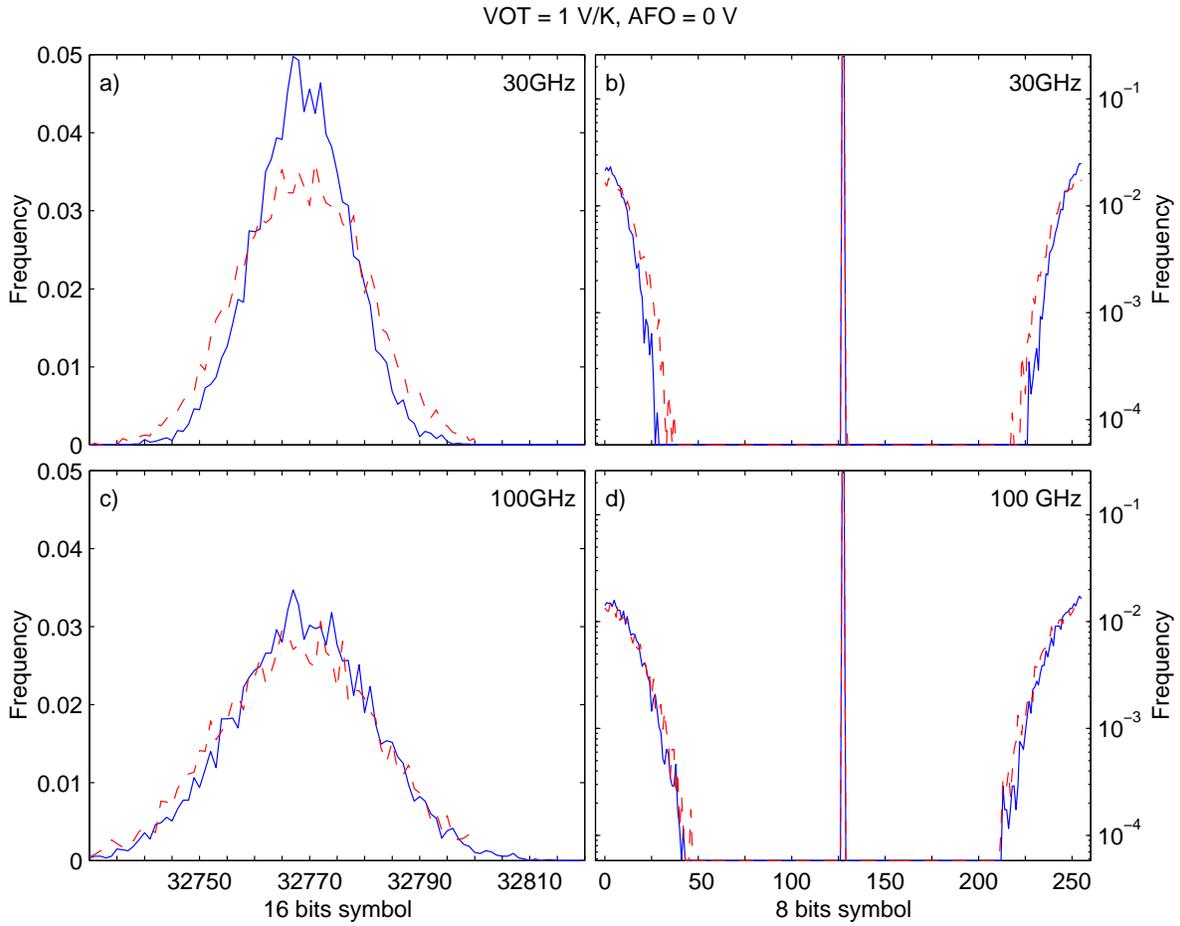}
 \caption[h]{
 Statistical distribution of 8 and 16 bits words for LFI simulated
 signals. Upper row is for 30 GHz, lower row for 100 GHz. Left
 column are the distributions of 16 bits words from the quantized
 signals, right column is for 8 bits words from the quantized
 signal, {\em full line} is the distribution for pure white noise,
 {\em dashed line} is the distribution for the full signal.
 }\label{fig:qstat}
\end{center} \end{figure}

\begin{figure}
\begin{center}
 \epsfysize=8cm \epsfbox{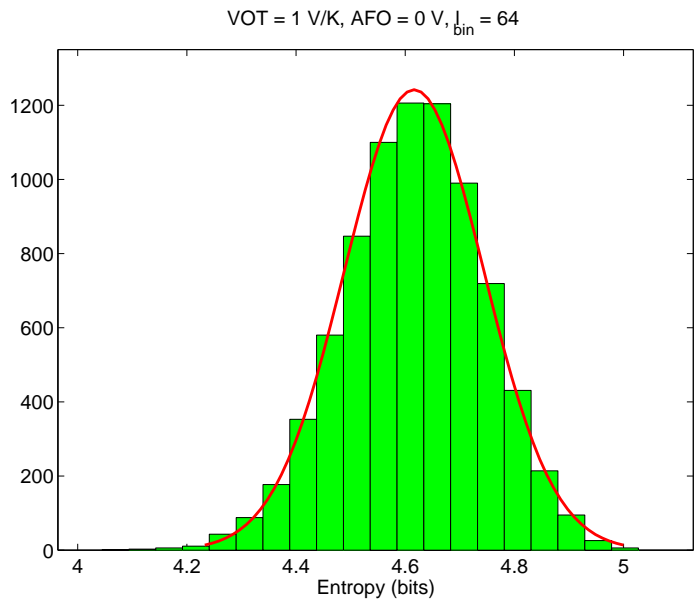}
 \caption{
  The entropy
  distribution per bunch for $l_{\mathrm{chunk}} = 64$\ samples, for
  the full signal at 30 GHz.
  } \label{fig:entropy:distribution}
\end{center}
\end{figure}

\begin{figure}[t]\begin{center}
 \epsfysize=10cm \epsfbox{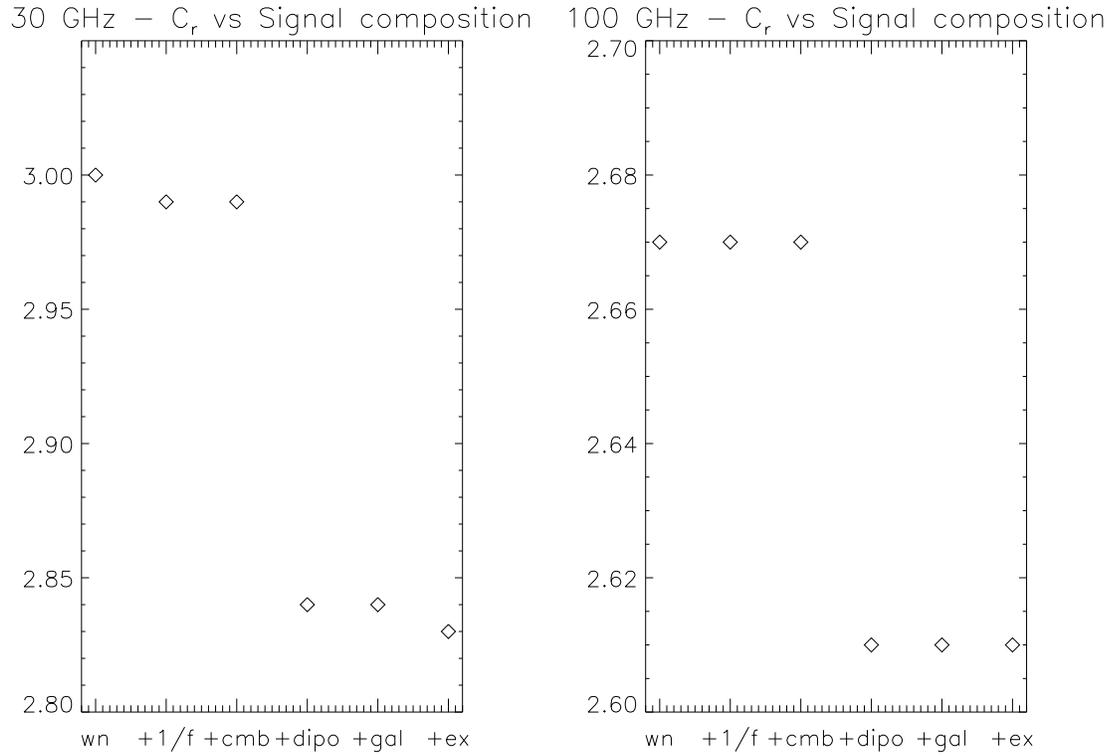}
 \caption[h]{
 Figures of merit for the arithmetic compression of
 order 1 ({\tt arith-n1}) for 30 GHz and 100 GHz channels.  Here
 $\AFO = 0$\ V, $\VOT = 1.0$\ V/K, $\Nc = 1$. The compression
 efficiency is plotted as a function of the incremental complexity
 of the signal composition: {\tt wn}\ means white noise only, {\tt
 +1/f}\ plus 1/f,  {\tt +cmb} plus CMB, {\tt +dipo} plus dipole,
 {\tt +gal} plus galaxy, {\tt +ex } plus extragalactic sources.
}
\label{fig:merith}
\end{center} \end{figure}

\begin{figure}[t]\begin{center}
 \epsfysize=10cm \epsfbox{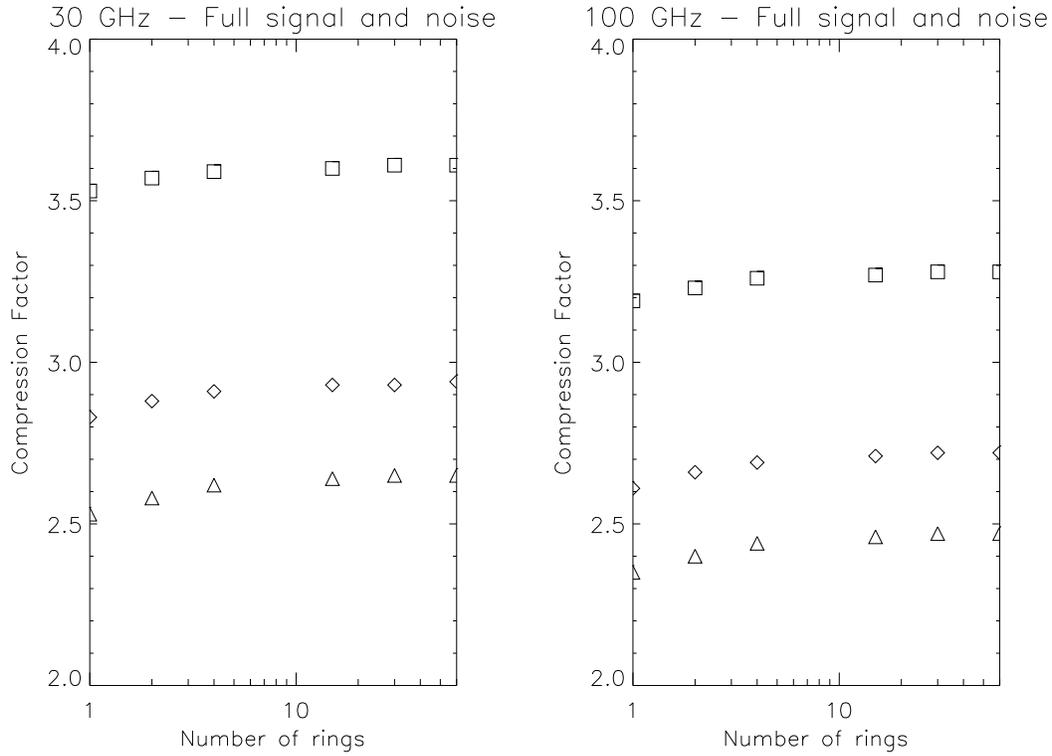} 
 \caption[h]{
 Compression rates for {\tt arith-n1}\ as a function of
 the \VOT\ and \Nc\ for a full simulated signal ({\tt wn} + {\tt
 1/f} + {\tt dp} + {\tt cmb} + {\tt dipo} + {\tt gal} + {\tt ex})
 (see also figure \ref{fig:merith} for details). From top to
 bottom:
 {\em Squares}: $\VOT = 1.5$\ V/K,
 {\em Diamonds}: $\VOT = 1.0$\ V/K,
 {\em Triangles}: $\VOT = 0.5$\ V/K.
}\label{fig:cr:ring}
\end{center} \end{figure}

\end{document}